\documentclass[fleqn,10pt]{wlscirep}
\usepackage[utf8]{inputenc}
\usepackage[T1]{fontenc}
\usepackage{braket}
\usepackage{subcaption}
\usepackage{graphicx}
\usepackage{amsthm}
\usepackage{placeins}
\usepackage{appendix}
\usepackage{tabularx}
\usepackage{makecell}
\usepackage{physics}
\usepackage{enumitem}
\usepackage[singlelinecheck=false]{caption}
\usepackage[font=small,skip=5pt]{caption}

\DeclareMathAlphabet{\mathcal}{OMS}{cmsy}{m}{n}
\SetMathAlphabet{\mathcal}{bold}{OMS}{cmsy}{b}{n}

\newtheorem{definition}{Definition}[section]

\title{Surrogate-guided optimization in quantum networks}

\author[1,2,*]{Luise Prielinger}
\author[1,2]{Álvaro G. Iñesta}
\author[1,2,3]{Gayane Vardoyan}
\affil[1]{QuTech, Delft University of Technolgy, Delft, The Netherlands.}
\affil[2]{EEMCS, Delft University of Technology, Delft, The Netherlands. }
\affil[3]{CICS, University of Massachusetts, Amherst, USA.}

\affil[*]{l.p.prielinger@tudelft.nl}

\keywords{Quantum Networks, Simulation Optimization, Surrogate Model, Support Vector Regression, Bandit Optimization}

\begin{abstract}
We propose an optimization algorithm to improve the design and performance of quantum communication networks.
When physical architectures become too complex for analytical methods, numerical simulation becomes essential to study quantum network behavior. Although highly informative, these simulations involve complex numerical functions without known analytical forms, making traditional optimization techniques that assume continuity, differentiability, or convexity inapplicable. Additionally, quantum network simulations are computationally demanding, rendering global approaches like Simulated Annealing or genetic algorithms,
 which require extensive function evaluations, impractical. We introduce a more efficient optimization workflow using machine learning models, which serve as surrogates for a given objective function. 
We demonstrate the effectiveness of our approach by applying it to three well-known optimization problems in quantum networking: quantum memory allocation for multiple network nodes, tuning an experimental parameter in all physical links of a quantum entanglement switch, and finding efficient protocol settings within a large asymmetric quantum network. The solutions found by our algorithm consistently outperform those obtained with our baseline approaches -- Simulated Annealing and Bayesian optimization -- in the allotted time limit by up to 18\% and 20\%, respectively. Our framework thus allows for more comprehensive quantum network studies, integrating surrogate-assisted optimization with existing quantum network simulators.
\end{abstract}

\begin{document}
\flushbottom
\maketitle
\thispagestyle{empty}

\section*{Introduction} \label{sec:intro}
Quantum network infrastructure has the potential to be integrated with today's Internet, serving entangled states to users who request them~\cite{wehner2018quantum}. These networks will enable a number of applications provably beyond the capabilities of classical technologies alone. Examples include verifiably secure communication~\cite{bennett2014quantum, ekert1991quantum}, advanced sensing tasks~\cite{giovannetti2004quantum, jozsa2000quantum} and blind quantum computation~\cite{broadbent2009universal, fitzsimons2017unconditionally}, among others. For quantum networks to achieve their intended potential, diverse solutions for all layers of the system stack have been put forward in recent years. At the physical layer, for example, so-called quantum repeaters have been proposed to assist with quantum information transport over long distances (see e.g., ref.~\cite{azuma2023quantum} for an in-depth review on quantum repeaters). The primary task at the software layer is to create efficient protocols that coordinate and control the intricate physical processes within the quantum network system stack~\cite{dahlberg2019link}. To this end, both software and hardware components have been extensively studied~\cite{van2008system, vardoyan2019stochastic, inesta2022optimal, vardoyan2023quantum, liao2022benchmarking, avis2023requirements, ghaderibaneh2022pre, kozlowski2020designing, da2021optimizing, vandam2024hardwarerequirementstrappedionbased} using analytical tools as well as numerical simulation to find feasible architectures and bring quantum network technology a step closer to the real world. These efforts include the use of optimization and machine learning techniques, which have been used, for example, to design new entanglement distribution protocols~\cite{Wallnöfer_2020, khatri2021policies, Haldar2024Fast}. While these studies are greatly informative, they generally assess only small-sized networks or operate under simplified assumptions, such as ideal hardware models or highly symmetric network typologies.

In this study, our primary goal is to utilize a surrogate-assisted optimization workflow to discover effective protocol and hardware parameter values under realistic conditions. To this end, we extend previous efforts and integrate comprehensive numerical simulators, such as NetSquid~\cite{coopmans2021netsquid} and SeQUeNCe~\cite{wu2021sequence} in our optimization framework. In essence,  a surrogate model~\cite{box1987empirical, queipo2005surrogate} is an approximation of a given objective function, which is built using data obtained from evaluations of the actual objective function. Importantly, instead of directly optimizing the computationally expensive objective function -- a common approach in global optimization techniques such as Simulated Annealing or genetic algorithms (see ref.~\cite{kochenderfer2019algorithms} for a review of numerical optimization methods) -- we employ a surrogate model to guide the iterative optimization process.

Common surrogate models like Gaussian processes\cite{Shahriari2016Taking} (used in Bayesian optimization) and neural networks\cite{tripathy2018deep} are based on complex theoretical frameworks. Gaussian processes, for example, depend significantly on the choice of a kernel function and they are known to be most effective in smaller, continuous search spaces with fewer than 20 variables\cite{frazier2018tutorial}. Neural networks, which consist of various specialized layers\cite{weiss1991computer}, require substantial data for optimal performance. Additionally, the outputs from these complex models are often more difficult to interpret compared to those from simpler models\cite{carvalho2019machine}. As a consequence, we opted for two well-known but less complex models --  Support Vector Regression~\cite{gunn1998support} (SVR) and Random Forest Regression~\cite{breiman2001random} (RF) --  due to their explainability, computational efficiency, and straightforward evaluation. Surrogate models have been utilized since the late 80s and ever since applied to various scientific domains, such as chemical engineering~\cite{bhosekar2018advances, wang2024identifying} and materials science~\cite{kusne2020fly}. In quantum sensing, Bayesian optimization has been utilized for quantum detectors~\cite{popp2021bayesian} and in quantum networking to find minimum hardware requirements ~\cite{stevense2024numerical} in small repeater chains and to calibrate experimental parameters~\cite{cortes2022sample}. To the best of our knowledge, surrogates have not yet been applied to any software component of quantum communication. 

In this work we introduce a surrogate-optimization framework and apply it to three quantum network use cases simulated in NetSquid~\cite{coopmans2021netsquid}, SeQUeNCe~\cite{wu2021sequence} and OptimizingCD~\cite{inesta2023performance}, respectively. Our contributions are as follows:
\begin{itemize}[leftmargin=*]
 \setlength\itemsep{0em}
    \item \textbf{Optimization algorithm integrating scientific software:} Our framework utilizes detailed simulations of practical quantum network scenarios. This method allows us to avoid reliance on purely theoretical predictions, which are often specific to certain  architectural or software stack characteristics like node connectivity or expected protocol behavior\cite{wu2021sequence, Wallnöfer_2020, khammassi2021, vardoyan2023quantum}. Furthermore, whereas traditional techniques might need hundreds or even thousands of optimization cycles\cite{alcazar2024enhancing, da2021optimizing}, our approach shows substantial improvements within less than a hundred cycles. This efficiency allows us to integrate computationally intensive simulations -- which can take several minutes to execute -- into our optimization process.
    \item \textbf{Handling of many parameters:} The scenarios we test encompass up to 100 network parameters. For the largest variable set, reference methods perform comparably to random search, whereas our approach significantly outperforms both. This capability is largely due to the simplicity of the SVR and RF models utilized, which exhibit only linear time complexity with the number of variables\cite{scikit-learn}.
    \item \textbf{Addressing multiple objectives:} In a quantum network where different parties might have their individual quality-of-service goals, our analysis allows us to determine an empirical estimate of the Pareto frontier. Using this solution set, we can assess the effectiveness of parameter settings in the context of multi-objective optimization. 
    \item \textbf{Diverse applicability:} The selected use cases present well-known optimization challenges of distributing entangled states in quantum networks. The solutions we find introduce competitive candidates to existing solutions found in literature. Furthermore, we test the applicability of our approach on a range of different entanglement distribution protocols, involving on-demand and continuous-distribution protocols\cite{chakraborty2019distributed, inesta2023performance}. 
\end{itemize}
We compare our approach to a Bayesian optimizer developed by Meta\cite{balandat2020botorch}, a traditional global optimization algorithm termed Simulated Annealing\cite{kirkpatrick1983optimization}, and uniform random search as baselines. In the scenarios we investigate, surrogate optimization consistently outperforms the selected reference methods by up to 20\% within the allotted time limits. 

In the following sections, we first introduce our surrogate-guided search for quantum network configurations, including the necessary preliminaries in Section \ref{sec:results}. In Section \ref{sec:usecases} we demonstrate the usability of our approach in various quantum network setups and detail its relation to previous work in Section \ref{sec:relwork}. Finally, Section \ref{sec:methods} describes our simulation experiments and the optimization methods used for comparison.

\section{Results} \label{sec:results}
\subsection{Preliminaries} \label{sec:prelim}
Consider a vector function $f( \mathbf{x}):X \rightarrow \mathbb{R}^{m_{f}}$ of arbitrary size $m_f \in \mathbb{N}$ describing a performance metric of a quantum network (e.g., the average fidelity of entangled states provided to a pair of nodes). The input to this function $\mathbf{x} = \{x_1, x_2, \ldots, x_N\}$ represents a set of quantum network parameter values, which may include both fixed (non-configurable) as well as tunable features. A parameter $x_p$ of a network configuration $\mathbf{x}$ is defined on a domain $X_p$. Together, the parameter domains form the input space: $X = X_1 \times X_2 \times \cdots \times X_N.$
We do not restrict $x_p$ to a specific datatype. For continuous and discrete values, $x_p$ is confined within some minimum-maximum bounds, $x_p^{\mathrm{min}}$ and $x_p^{\mathrm{max}}$, respectively, while ordinal and categorical parameters are represented as value sets. 

Based on $f$, we can formulate an objective function $U(f,\mathbf{x})$ reflecting the way in which a quantum network perceives \emph{utility}\cite{vardoyan2023quantum}. Essentially, while $f(\mathbf{x})$ describes some output of the network, the function $U(f, \mathbf{x}):\mathbb{R}^{m_f} \times X \rightarrow \mathbb{R}^{m}$, $m\in\mathbb{N}$, assesses how much utility can be derived from this output. Although $U$ can in principle represent any general objective, in our analyzed use cases, one element of $U$, denoted $U^{(i)}$, consistently represents the utility perceived by a network user $i$. We will explore some relevant examples of utility functions $U$ based on distillable entanglement~\cite{rains2001semidefinite}, request completions~\cite{wu2021sequence}, and virtual connectivity~\cite{inesta2023performance}.  

In a quantum network, many processes are inherently probabilistic. For example, creating entanglement between nodes often requires multiple attempts following a geometric distribution~\cite{azuma2023quantum}. As a consequence, we assume $f$ and thus also the utility $U(f, \mathbf{x})$ to be \emph{stochastic} functions. We denote the utility perceived by user $i$ with a random variable $Y^{(i)} \sim U^{(i)}(f, \mathbf{x})$ with expectation value $E[Y^{(i)}]$.  As $f(\mathbf{x})$ is evaluated via numerical simulation, we retrieve a sample $\{ U^{(i)}_1(f, \mathbf{x}), U^{(i)}_2(f, \mathbf{x}), \dots, U^{(i)}_n(f, \mathbf{x})\}$ via $n$ simulation runs to estimate the expected behaviour $E[Y^{(i)}]$ for each user $i$ with the sample mean $\Bar{U}^{(i)}(f, \mathbf{x})=\frac{1}{n} \sum_{j=1}^n U^{(i)}_{j}(f, \mathbf{x})$. In this work, our optimization objective is to maximize the aggregate utility of the network, i.e., the sum of all individual user utilities.

\begin{definition}[Utility maximization over configurable variables] \label{def:utilitymax}
We aim to maximize the aggregate utility over all users by configuring parameter values within the tunable portion of the domain $X$, $X_{\mathrm{conf}}\subseteq X$. This results in the objective
$\max_\mathbf{s} \ U_{aggr}(f, \mathbf{s}) := \sum_{i=1}^{m} \Bar{U}^{(i)}(f, \mathbf{s}),$ where $\mathbf{s}\in X_{\mathrm{conf}}.$ 
Here, the set $X \setminus X_{\mathrm{conf}}$ denotes all parameters which are fixed.
\end{definition}

\subsection{Surrogate-assisted Search} \label{sec:workflow}

\begin{figure}[ht]
    \centering\includegraphics[width=0.95\linewidth]{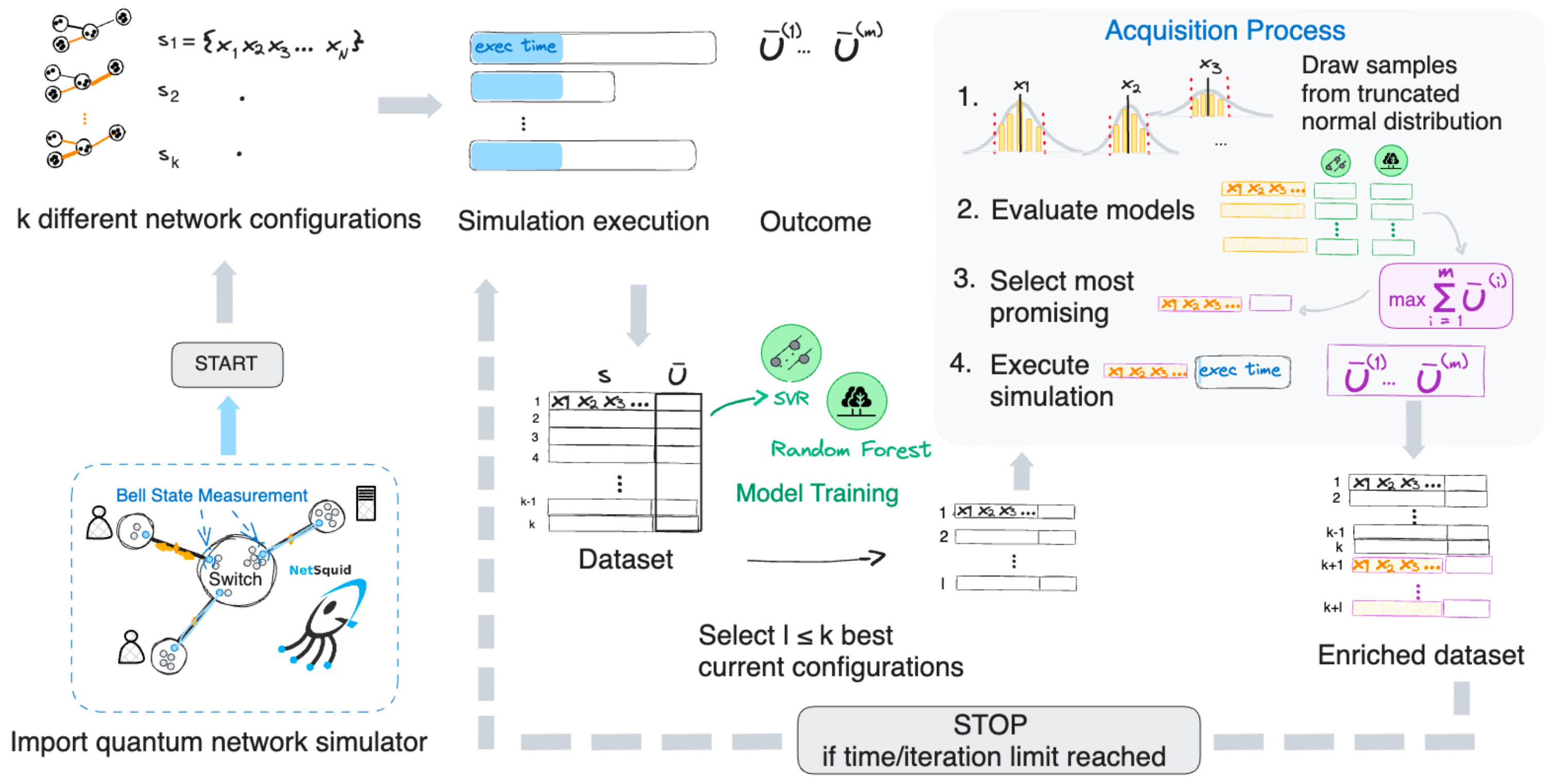}
    \caption{Simplified surrogate-assisted workflow. After importing a quantum network simulation, the algorithm generates $k_0$ different network configurations $\{\mathbf{s}_1, \dots, \mathbf{s}_{k_0}\}$ from the search domain. Execution of the quantum network simulation using the latter yields the initial training data of parameter sets $\{\mathbf{s}_i\}$ along with their means $[\Bar{U}^{(1)}(f,\mathbf{s}_i), \cdots, \Bar{U}^{(m)}(f,\mathbf{s}_i)]$. Machine learning models train on the dataset and their performance is evaluated. In the acquisition process, the best model in the cycle predicts utility values for sampled configurations. Then, the parameter set  associated with the highest predicted utility is passed to the simulation, executed, and the outcome appended to the training data. This cycle repeats until the optimization concludes upon reaching a wall-clock time or number of cycles $T$.} 
    \label{fig:workflow}
\end{figure}

We present our optimization workflow depicted in Figure \ref{fig:workflow}, which progresses through successive cycles $t \in \{0, \dots, T\}$. It begins by randomly generating $k_{0}$ initial input sets $\{\mathbf{s}_1, \mathbf{s}_2, \dots, \mathbf{s}_{k_0}\}$ from the search domain $X_\mathrm{conf}$, where each set is a unique configuration of parameter values. Then, for each input set $\mathbf{s}_i$, a quantum network simulation is run $n$ times to produce mean utility outputs $[\Bar{U}^{(1)}(f, \mathbf{s}_i), \cdots, \Bar{U}^{(m)}(f, \mathbf{s}_i)]$. Each of the $n$ runs lasts $T_{\mathrm{sim}}$ simulation time units. The input configurations together with the associated utility outputs form the initial \emph{dataset}, concluding the first cycle, $t=0$. 
At the beginning of the next cycle, $t=1$, the two machine learning models (SVR, RF) are evaluated using five-fold cross validation; thereby the models are trained on 80\% of the dataset (the training set) and evaluated on the remaining data (the test set). Their performance is measured using the mean absolute error between predicted and actual values in the test set. For further details on the model training refer to Supplementary Notes 1 and 2.

In the subsequent \emph{acquisition process}, the currently better performing machine learning model is used to select new configurations. To this end, \emph{for each} of $l$ so far best performing configurations $\{\mathbf{s}^\mathrm{top}_1, \dots, \mathbf{s}^\mathrm{top}_l\}$, a promising neighbor is identified: first, a number of points near a configuration $\mathbf{s}^\mathrm{top}_i$ is sampled from truncated normal distributions centered around each parameter value in $\mathbf{s}^\mathrm{top}_i$. Formally, parameter values are sampled from $\mathcal{N}_{\mathrm{trunc}}(\mu_{p}, \sigma_{p}(t,d))$. These distributions have mean $\mu_{p}=x_p$, standard deviation $\sigma_{p}(t,d) = \gamma(t,d) (x^\mathrm{max}_p - x^\mathrm{min}_p)/2$, and are truncated at the bounds $x^\mathrm{min}_p$ and $x^\mathrm{max}_p$ (see Section \ref{sec:prelim}). We gradually shift from exploration towards exploitation by progressively narrowing the standard deviation using a monotonically-decreasing transition function $\gamma(t,d)$ depending on the elapsed cycles $t$ and an exploitation degree $d \geq 1$. This adjustment ensures that as each cycle progresses, the focus on exploring unknown points in $X_{\mathrm{conf}}$ turns to exploiting known good areas of the search space (see Supplementary Notes 1 for further details on the exploitation strategy). Furthermore, the number of sampling points increases incrementally with each cycle $t$. This approach allows for less costly computation during early cycles, when the models' performance is lower, and progressively dedicating more time as the models improve and accumulate knowledge (see details in Supplementary Note 2). 

Once the points are passed to a model, it predicts utility values based on its current knowledge, and the configuration associated with the \emph{highest} utility is returned. This leads to $l$ new configurations $\{\mathbf{s}^\mathrm{next}_1, ..., \mathbf{s}^\mathrm{next}_l\}$ to be evaluated by the simulation. Finally the newly generated parameter sets and associated simulation outputs are added to the dataset, $k_{1} = k_0 + l$, which completes the optimization cycle $t=1$. This cycle repeats until a time limit or maximum number of iterations, $t=T$ is reached. By default, $k_0$ and $l$ are chosen according to the number of compute resources available for parallel execution. Table \ref{tab:surparams} summarizes the parameters used.

\begin{table}[ht!]
  \begin{tabular}{|cl|}
    \hline
    Symbol & \textbf{Description} \\ 
    \hline
    $T$ & Optimization limit, can be specified as wall-clock time limit   \\
    & or maximum number of optimization cycles\\
    $t$ & Elapsed time $t\in [0,T]$ or cycle $t\in\{0,\dots,T\}$\\
    $n$ & Number of simulation evaluations used to compute $\bar{U}$\\
    $l$ & Number of points explored in an optimization cycle\\
    $d$&  Exploitation degree (see Supplementary Note 1)\\
    $k_t$& Number of points in the dataset at cycle $t$, where $k_{t+1} = k_t + l$ and $k_0 = l$\\
    $T_{\mathrm{sim}}$ & Simulation time of the quantum network model (specific to use case)\\
    
    \hline
  \end{tabular}
  \caption{Parameters used in surrogate-assisted search.}
  \label{tab:surparams}
\end{table}
 Every algorithm comes with its limitations: heuristic approaches, like the one described, do not guarantee a globally optimal solution but instead provide approximations once a termination criterion is met~\cite{kochenderfer2019algorithms}. This criterion is primarily dictated by the available computational resources. For example, considering the computationally demanding nature of our simulation functions, we set the termination criterion  to not exceed $T=100$ optimization cycles. Further, we must rely on seeds to guarantee reproducibility as both the simulation as well as the machine learning models used in the acquisition process are inherently stochastic. In addition, selecting settings for workflow parameters like the degree of exploitation $d$ is crucial and requires understanding their specific function (see Supplementary Note 1). Finally, the numerical effort to train and evaluate machine learning models makes our approach naturally less suitable for problem instances that are analytically tractable. 

\section{Use Cases}\label{sec:usecases}
\subsection{A Quantum Entanglement Switch} \label{sec:QES}
\begin{figure}[ht]
    \centering\includegraphics[width=0.9\linewidth]{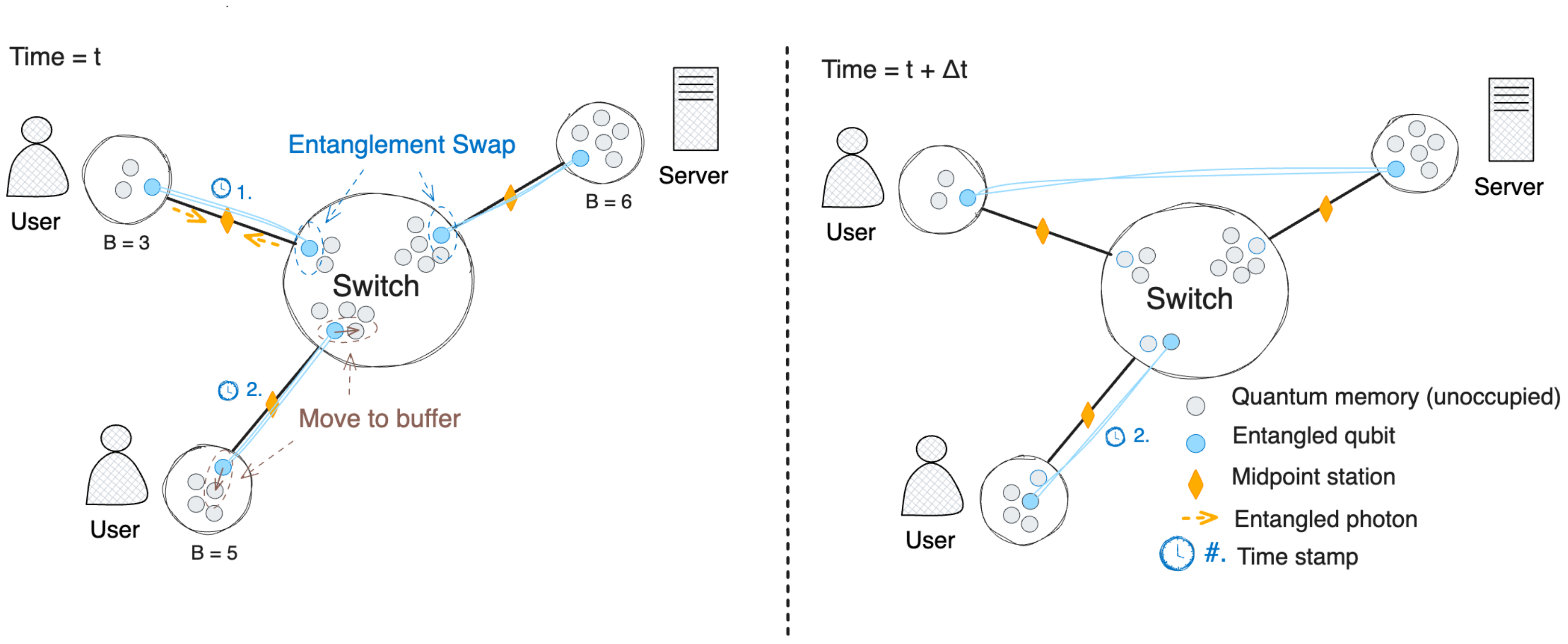}
    \caption{Quantum Entanglement Switch (QES) serving two users and a server. Each fiber link is equipped with a midpoint station (orange diamond) that measures photons entangled with communication qubits located at nodes. After an entangled state is generated, it is immediately transferred to a free memory qubit in the buffer (shown with brown arrows) along with a time stamp (small blue clocks). Should the buffer be full, the oldest state is discarded and the memory receives the fresh link. The switch can only execute one entanglement swap at a time. By default, the user who holds the oldest link (smallest time stamp) is given priority for the swap.}
    \label{fig:QES-sketch}
\end{figure}

A quantum entanglement switch (QES) is a canonical example of a simple quantum network\cite{vardoyan2019stochastic, vardoyan2023capacity, nain2020analysis, vardoyan2023quantum, coopmans2021netsquid} where a set of user nodes are connected to a central hub that distributes entanglement among them. We assume that each node is equipped with memory qubits and is connected to the QES via a fiber optic link to generate so-called link-level entangled states, henceforth referred to simply as \emph{links}. In our setup, one node is designated as a server, while the remaining nodes operate as users, see Figure \ref{fig:QES-sketch}. The QES aims to create a direct entangled connection between the server and each of the users, which we term \emph{end-to-end link}, in the form of bipartite entanglement.
The QES accomplishes this
using a process called entanglement swapping\cite{azuma2023quantum}, 
during which quantum operations at the switch transform two links into an end-to-end link between a user and the server. After generation, the state is transferred to an available memory qubit in the buffer of size $B$. When the buffer is full, the oldest state is discarded.

We assume that link-level entanglement generation uses the single-click scheme~\cite{humphreys2018deterministic}, wherein each physical link $l$ has a midpoint station with a beamsplitter and two photon detectors. Adjacent nodes each contain a communication qubit entangled with an emitted photon. When these photons arrive at the midpoint station, they are measured to generate an entangled state between the communication qubits. Under high photon losses, this scheme shows a tradeoff between entanglement generation rate and fidelity due to the bright-state population parameter $\alpha_l$. A higher $\alpha_l$ increases photon emission but at the same time degrades entanglement fidelity. Consequently, the success probability $p_{gen,l}$ of entanglement generation on link $l$ grows with $\alpha_l$, while the fidelity $F_l$ decreases as $F_l = 1-\alpha_l$. 
We approximate this physical behaviour at each physical link using NetSquid's depolarizing-error model. A depolarizing error is a type of quantum error that, with a certain probability $p_\mathrm{depol,l}\propto \alpha_l$, transforms the photon's polarization state into a completely mixed state. For further details on how we model the physical quantum states based on ref.\cite{vardoyan2023quantum} and ref.\cite{coopmans2021netsquid} we refer the reader to Supplementary Note 3. 

Our goal is to balance the rate-fidelity tradeoff across all $N$ physical links $\mathbf{s}_\alpha = \{\alpha_1, \alpha_2, \dots, \alpha_N\}$, in order to maximize utility served to all users. Following the approach of ref.\cite{vardoyan2023quantum}, utility is assessed based on distillable entanglement, which quantifies how many degraded entangled pairs can be restored to a state useful for quantum communication tasks.
\begin{definition}[Utility Based on Distillable Entanglement] \label{def:qes-obj}
A user's utility $U^{(i)}$ is based on the average end-to-end entanglement rate $R^{(i)}$ and fidelity $F^{(i)}$ of end-to-end states established with the server. To calculate a user's utility, the fidelity $F^{(i)}$ is passed to the  so-called "yield of the hashing protocol" $D_H$ which serves as a lower bound on distillable entanglement\cite{bennett1996mixed}:
    \begin{align}
    U^{(i)}( \mathbf{s_\alpha}) :=\log\left(R^{(i)}( \mathbf{s_{\alpha}})\cdot D_H(F^{(i)}(\mathbf{s_{\alpha}}))\right) \text{, with}\ \ 
        D_H(F) := \max\left(1 + F \log_2 F + (1-F) \log_2 \frac{1-F}{3}, 0\right). \label{eq:end-to-endraw}
    \end{align}
    \end{definition}
Using this definition, we can write our objective to maximize the sum of all user utility values $U^{(i)}$ as
\begin{align}
    \max_{\mathbf{s}_\alpha}\ U_{aggr}(\mathbf{s}_\alpha) = \sum_{i=1}^{m}  U^{(i)}(\mathbf{s}_\alpha),
\end{align}
where $m$ denotes the number of users. 

Although this problem can be solved analytically for a small number of users \cite{vardoyan2023quantum}, the objective function can be non-convex, complicating the application of analytical or even algorithmic optimization techniques, such as gradient ascent. Using simulation in combination with surrogate optimization might thus provide a meaningful solution while mitigating these difficulties. We approximate average end-to-end fidelities $\Bar{F}^{(i)}(f,\mathbf{s_{\alpha}})$ as well as average rates $\bar{R}^{(i)}(f,\mathbf{s_{\alpha}})$ using $n$ runs of our simulation $f(\mathbf{s_{\alpha}})$ implemented with NetSquid. We then use these quantities to compute sample means of user utilities: $\Bar{U}^{(i)}(f, \mathbf{s_{\alpha}}) = \frac{1}{n} \sum^n_{i=1} U^{(i)}(f, \mathbf{s_{\alpha}})$.

First, we apply our surrogate workflow to a small QES network serving two users and a server; we thereby recover a selection of results from the analytical study in ref.~\cite{vardoyan2023quantum}, described in Supplementary Note 3.  

\begin{figure}[ht!]
\centering
\begin{minipage}{.55\textwidth}
\centering
\includegraphics[width=\linewidth]{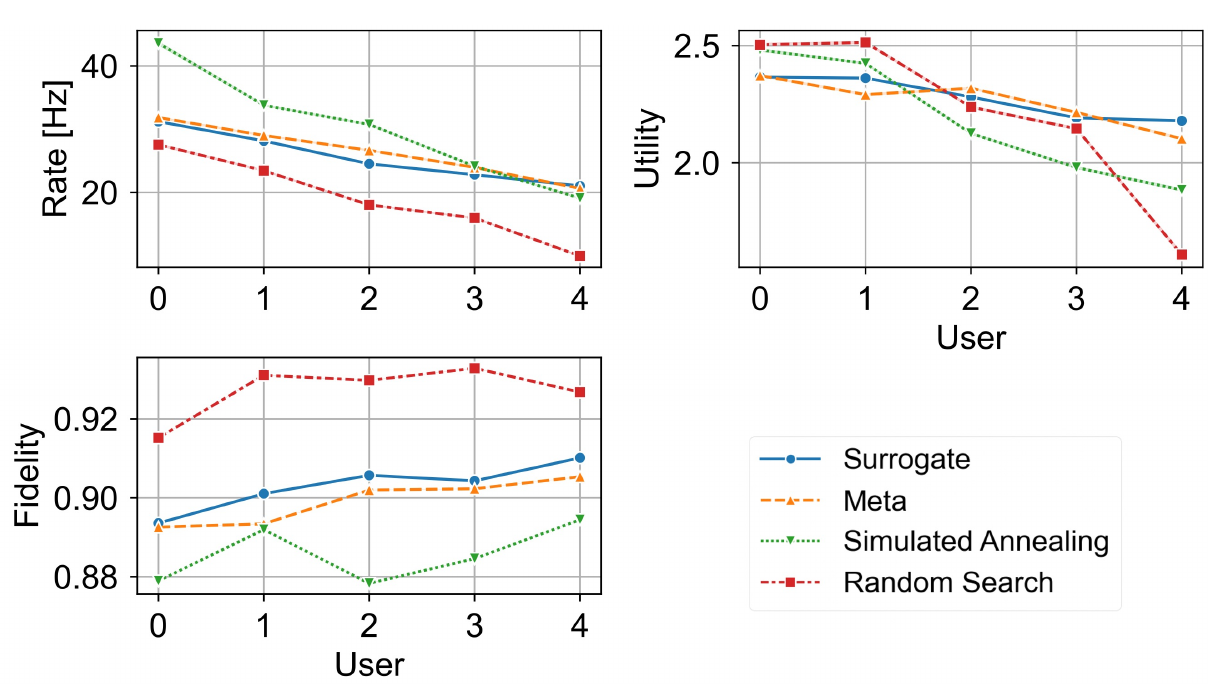}
\subcaption{\centering Performance at the user level.}
\end{minipage}
\begin{minipage}{.4\textwidth}
\centering
\includegraphics[width=0.8\linewidth]{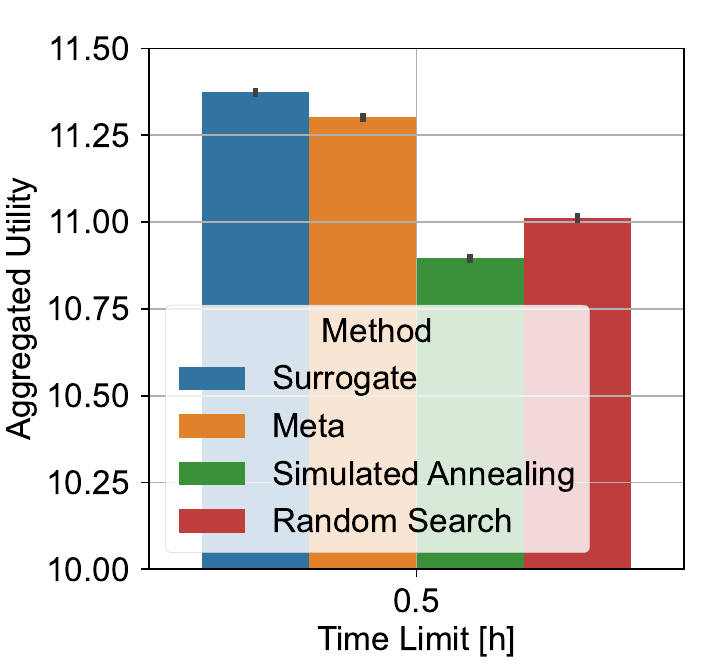}
\subcaption{\centering Best found aggregate utility values.}
\end{minipage}
    \caption{Utility maximization results for five users at varying distances from the switch. \textbf{(a)} The average rate and fidelity at which each user receives end-to-end links with the server, as well as the resulting utility values based on distillable entanglement. Each marker presents the mean (and standard error) of $n_{exec}=10^3$ simulation evaluations (standard error is smaller than marker size). \textbf{(b)} Aggregate utility over all users. Parameters used in the surrogate workflow: exploitation degree $d=4$, maximum execution time $T=30$ min, $n=20$ simulation runs.}
    \label{fig:usecase1-example2}
\end{figure}

Next, we extend the setup to one involving five users located at different distances from the switch: 10, 20, 30, 40, and 50 km. The server is located close to the switch, at a fixed distance of 5 km.   We conduct a comparison to reference methods (Section \ref{sec:methods-baselines}) involving ten independent repetitions for each optimization method. From these repetitions, we select the solution that exhibits the best performance according to the objective function. A subsequent simulation of these solutions is then carried out with $n_{exec} = 10^3$ runs each. These additional runs ensure that the collected statistics (e.g., mean aggregate utility) are sufficiently representative of the distributions that give rise to them.  Figure \ref{fig:usecase1-example2}\textbf{a} depicts the resulting rate  $\Bar{F}^{(i)}$ as well as the average rate $\bar{R}^{(i)}$ of end-to-end links per user $i$. Simulated Annealing and random search tend to find solutions with either relatively high rates but low fidelity, or low rates but high fidelity for each user. In contrast, both Meta and surrogate optimization strike a more meaningful balance between rate and fidelity, resulting in higher aggregate utility, depicted in Figure~\ref{fig:usecase1-example2}\textbf{b}. Overall, we observe slightly higher utility outcomes of the surrogate workflow compared to the reference methods: $0.5$\%, $4.3$\%, and $3.3$\% for Meta, Simulated Annealing, and random search, respectively. 
As expected, Meta and surrogate optimization show similar performance. This similarity arises because Meta, as outlined in ref.~\cite{balandat2020botorch}, utilizes Bayesian optimization -- a method well-suited for problems involving only continuous variables and fewer than 20 variables\cite{frazier2018tutorial}.

While here our algorithm does not provide a drastic improvement compared to baselines, the goal of this use case is to show that our surrogate workflow enables a straightforward transition from a simple and highly symmetric configuration to a more complex setup, which may no longer be amenable to analytical or exact solutions. 
In the next use case, we investigate a metropolitan-area quantum network involving a more comprehensive system stack and nine discrete hardware parameters to configure; this leads to a significant divergence between the Meta algorithm and our surrogate approach.

\subsection{Memory Allocation in a Metropolitan Quantum Network} \label{sec:RB}
\begin{figure}[ht]
\centering
\begin{minipage}{.5\textwidth}
\includegraphics[width=\linewidth]{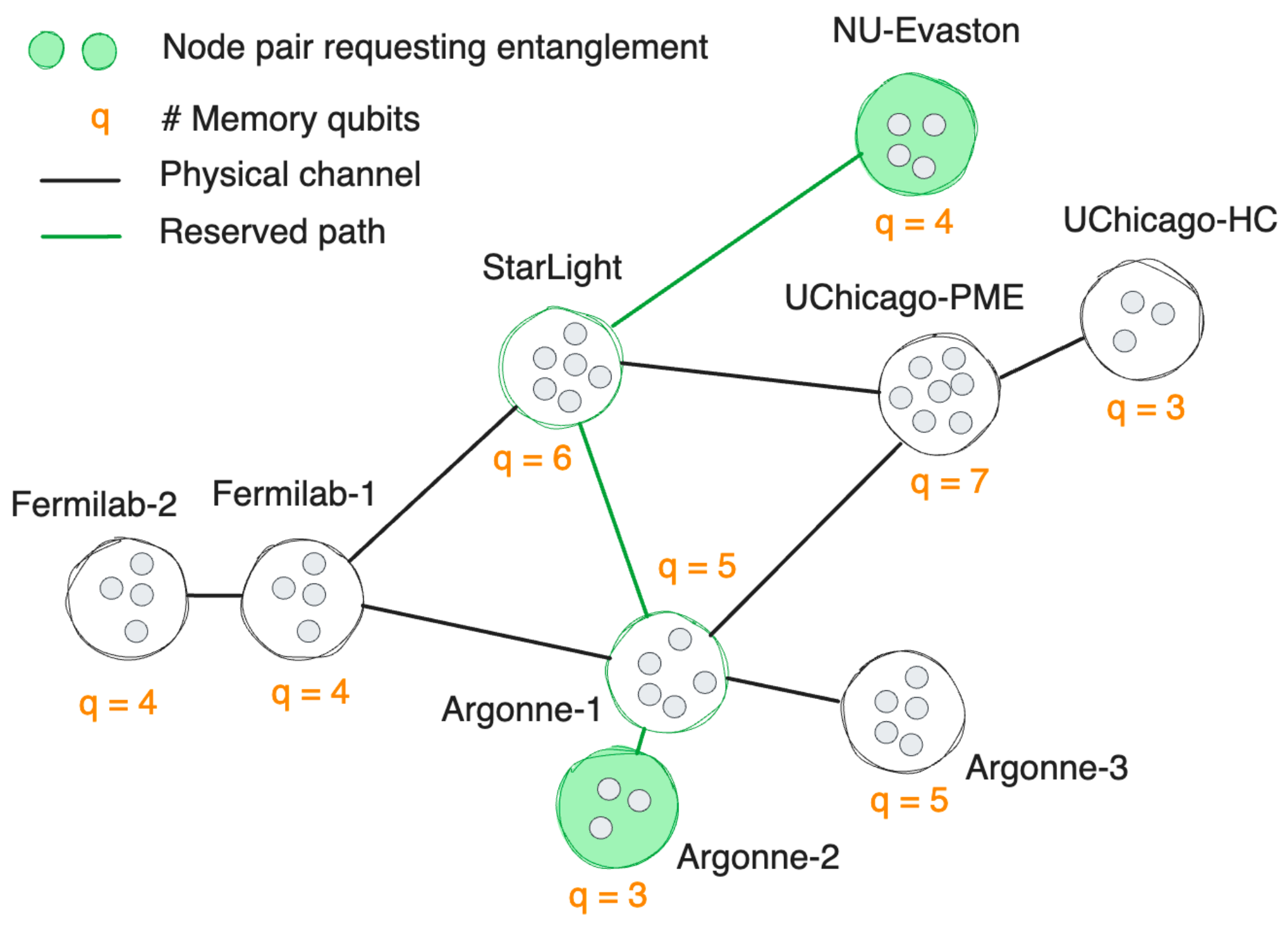}
\caption{Illustration of the metropolitan quantum network studied in ref.\cite{wu2021sequence}. Each node $i$ receives a number of memory qubits $q_i$ according to an allocation $\mathbf{q}$ (example values in orange). All nodes, one after the other, request to share entanglement with a randomly chosen other node. The network system stack fulfills the requests sequentially until the simulation time limit $T_\mathrm{sim}$ is reached.}
\label{fig:sequencenet}
\end{minipage}\hfill
\begin{minipage}{.46\textwidth}
    \centering
    \includegraphics[width=\linewidth]{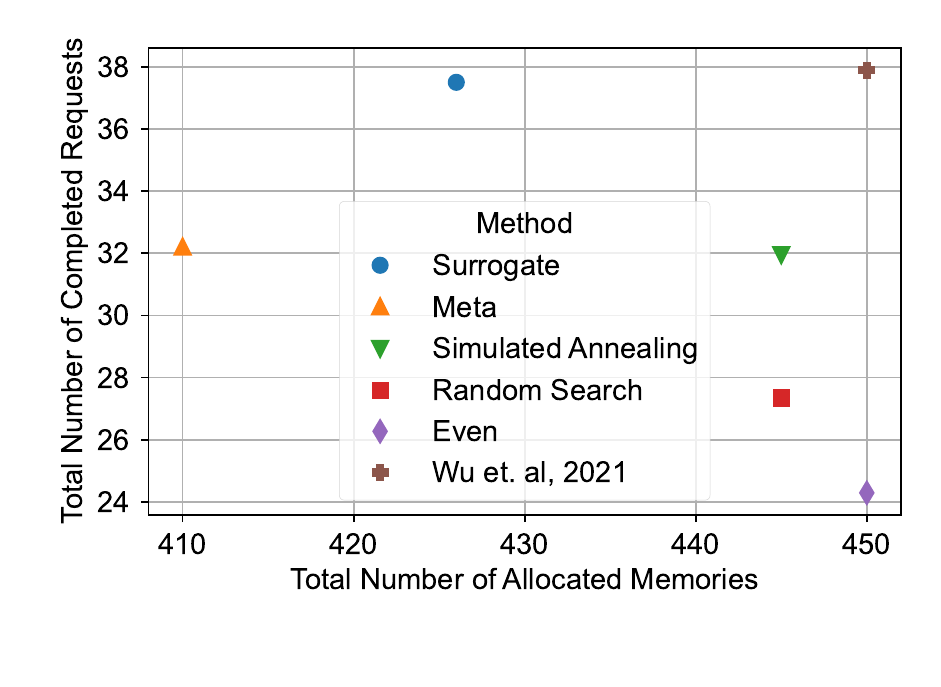}
    \caption{Sum of all completed requests over the total number of allocated quantum memories for solutions found by different methods. For the surrogate workflow, we use the following parameter settings: $T=25$ h, $d=6$, $n=5$. Each marker represents the average of $n_{exec}=10^3$ simulation runs with standard errors smaller than marker size. }
    \label{fig:usecase2-results}
\end{minipage}
\end{figure}

We study another previously investigated quantum network setup: a metropolitan-area network, depicted in Figure \ref{fig:sequencenet}. The network is modelled using the software package SeQUeNCe\cite{wu2021sequence}. All $m=9$ nodes of the quantum network are users, each possessing at least ten memory qubits $q_i\ge10$, which are modelled based on single erbium (Er) ions in crystal\cite{ranvcic2018coherence}. Nodes are equipped with a network manager for orchestrating local resource allocation and classical communication between nodes. Further, they are capable of generating entanglement with their immediate neighbors using dedicated midpoint stations at each optical link. Nodes also have the ability to perform entanglement swapping and  execute the BBPSSW purification protocol\cite{bennett1996purification} to enhance fidelity. In this scenario, utility is measured by the network’s ability to meet entanglement requests issued by user pairs. One user at a time selects another user uniformly at random and submits a request to its network manager. The network manager identifies the shortest route and announces the request to the nodes' network managers on the route. Each request specifies three key requirements: 1) the desired target fidelity of the end-to-end entangled state chosen uniformly at random between $0.8$ and $1$; 2) a number of memory qubits in all nodes along the path, selected uniformly at random 
from $\{10,\dots, \min(\frac{q_{\mathrm{init}}}{2}, q_{\mathrm{resp}})\}$, where $q_{\mathrm{init}}$ and $q_{\mathrm{resp}}$ are the number of qubits at the initiating and responding node, respectively,
and 3) a required period between one and two seconds to consume the entangled state. If a request fails (for example, due to insufficient end-to-end fidelity or insufficient number of memories) the request is reattempted using a different randomly selected number of memories until success. After the first request is completed, another user submits one. When all users have submitted and completed a request, the process starts anew. For a detailed explanation of the model used and its parameters, we direct the reader to the comprehensive description provided in ref.~\cite{wu2021sequence} and Supplementary Note 4. 

We simulate this process of submitting and fulfilling requests over $T_{\mathrm{sim}} = 20$ s.  Crucially, the number of requested memory qubits directly influences the potential for multiple rounds of entanglement purification\cite{dur2007entanglement}, thereby increasing the likelihood of achieving the desired fidelity. Thus, similar to ref.\cite{wu2021sequence}, we aim to maximize the number of requests the network can serve by distributing a given memory budget of $b=450$ memory qubits across the network's nodes.

\begin{definition}[Memory Allocation Problem] \label{def:budget} Let $b$ be a budget of memory qubits and $c_{i}$ the maximum number of memories node $i$ can hold. The problem is to determine a memory allocation  $\mathbf{q} = \{q_1, \dots, q_m\}$ across all $m$ nodes that maximizes the number of completed requests $U^{(i)}$ aggregated over all users $i \in \{1,\dots,m\}$. We define the following maximization problem:
\begin{align}
    &\max_{\mathbf{q}}\  \sum^m_{i=1} U^{(i)}(\mathbf{q}, b) - P(\mathbf{q}, b),\\
    &\text{where}\ P(\mathbf{q}, b):= \max(0, (\sum^m_{i=1} q_i)-b) \label{eq:constraint1} \quad \text{with}\quad0\le q_i\le c_{i} \quad \forall i \in \{1,\dots,m\},
\end{align}
such that for memories exceeding the budget $b$, the function $P\ge 0$ adds a scalar penalty.
\end{definition}

As a remark, Definition \ref{def:budget} is not unique, but one of several possible formulations. It presents a less restrictive optimization problem than the problem solved exhaustively in ref.\cite{wu2021sequence}. There, a fixed budget of 450 qubits in total is distributed across network nodes ($\sum^m_{i=1} q_i = b$), while we only induce a penalty if $\sum^m_{i=1} q_i > b$. Relaxing the fixed budget constraint to the penalty $P$ is a different problem statement than the one solved in ref.\cite{wu2021sequence}. However, comparison to solutions of the fixed-budget problem remains valuable, as found solutions can provide insight into the tradeoff between resource (quantum memory) usage and gain in utility.

We utilize the surrogate workflow as well as five reference methods (see Section \ref{sec:methods-baselines} for details), to find solutions to the memory allocation problem, where we compute $\Bar{U}^{(i)}(f(\mathbf{q}), b)$ using the adapted network simulation of ref.\cite{wu2021sequence} as $f(\mathbf{q})$. Note that the execution time depends on the input values and takes on average nine minutes on our system using $n=5$ simulation runs to calculate sample averages. This leads to long execution times, bringing about $T=25$ hours to be a reasonable time limit for the optimization process. The best found solution from ten independent trials per method is presented in Supplementary Table \ref{tab:usecase2-solutions}. Figure~\ref{fig:usecase2-results} displays the associated results: the average number of requests fulfilled, calculated from $n_{\mathrm{exec}}=10^3$ simulation runs, vs the total number of memories allocated. The solution found by surrogate optimization outperforms solutions found by Meta and Simulated Annealing by completing, on average, five additional requests (totaling 37.5). Further, it completes only 0.2 requests less when compared to the allocation derived from an exhaustive graph traversal algorithm, as detailed in ref.~\cite{wu2021sequence}. Notably, the surrogate optimizer's solution conserves 27 memory qubits compared to this exhaustive method, amounting to a 6\% resource saving with a minor performance decrease of only 0.5\%. The baseline from ref.~\cite{wu2021sequence}, employing the \emph{Even} allocation where each node possesses 50 memory qubits, exhibits the poorest performance. Using the Even allocation, the network completes less than $25$ requests, which is on average three requests worse than the solution found by the random search baseline.

With these results, we demonstrate that even though quantum network simulation can come with a heavy computational overhead, surrogates enable the discovery of solutions that are competitive to exhaustive search approaches, e.g., as used in ref.\cite{wu2021sequence}.

\subsection{Continuous Distribution of Entanglement} \label{sec:CD}
\begin{figure}[hb!]
\centering
\begin{minipage}{.4\textwidth}
    \centering
    \includegraphics[width=\linewidth]{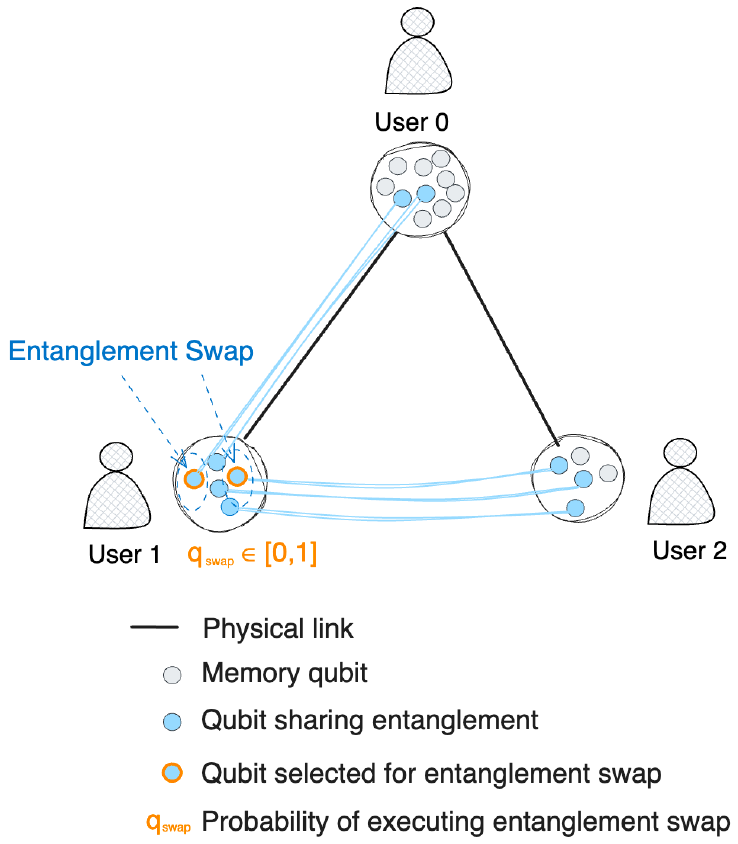}
\subcaption{\centering Three-user network.}
\end{minipage}
\begin{minipage}{.52\textwidth}
    \centering
    \includegraphics[width=\linewidth]{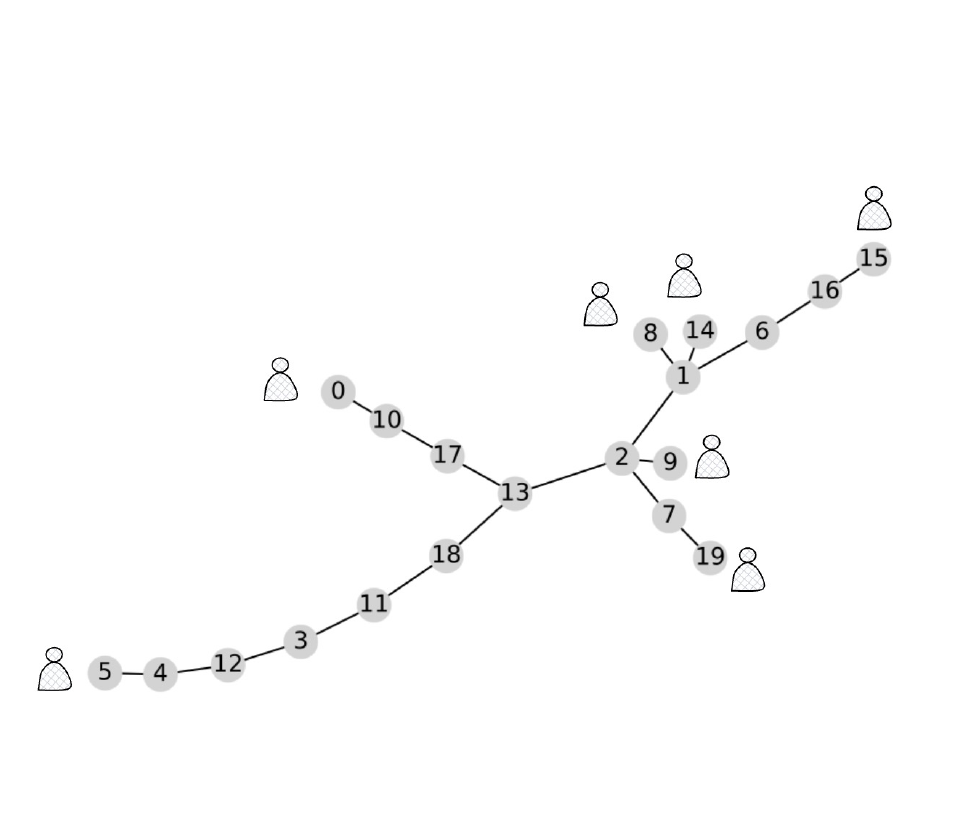}
\subcaption{\centering 20-node network serving seven users.}
\end{minipage}
\begin{minipage}{.9\textwidth}
    \centering
    \includegraphics[width=\linewidth]{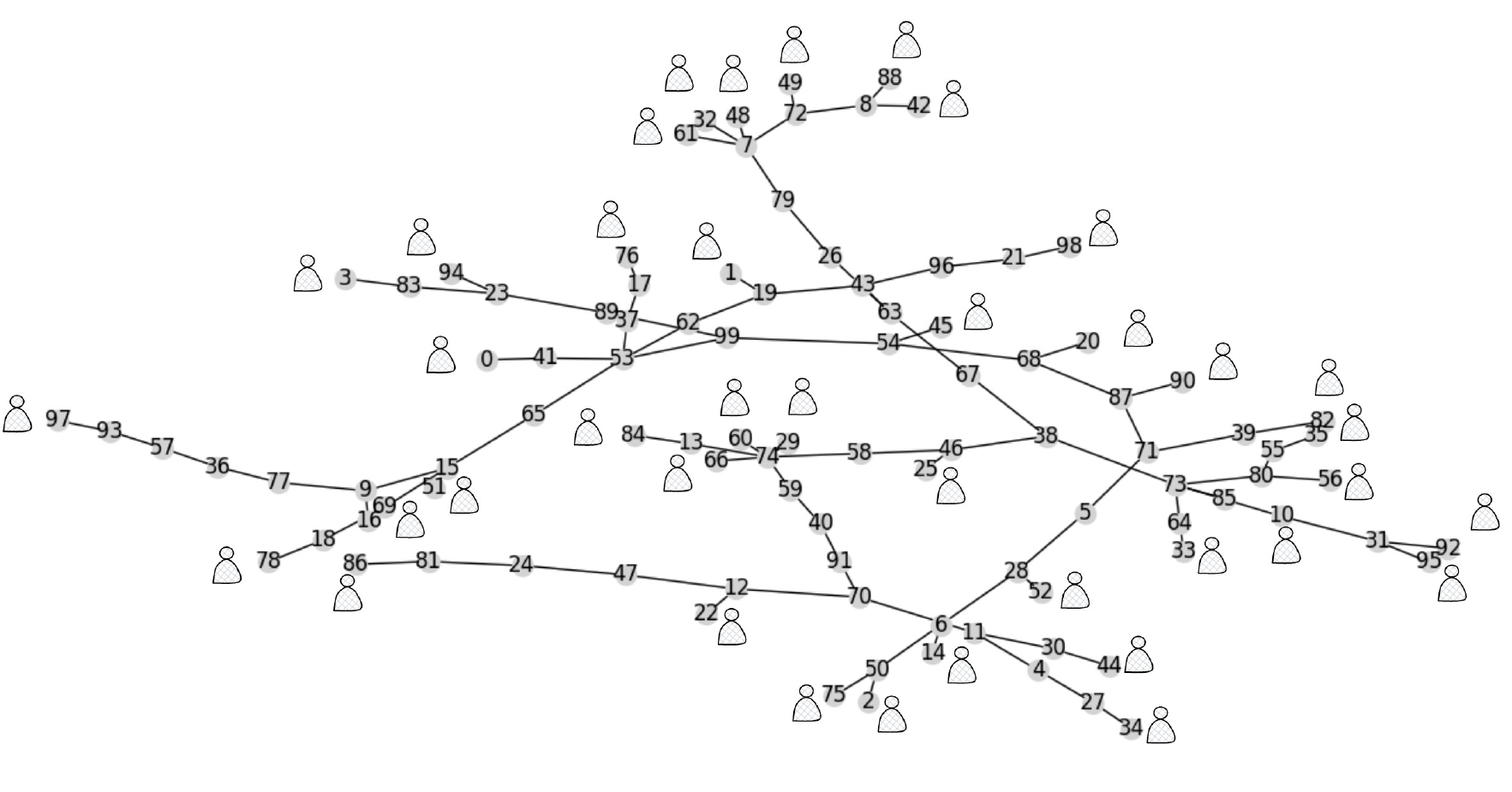}
\subcaption{\centering 100-node network serving 39 users.}
\end{minipage}
    \caption{Quantum network topologies considered in continuous-distribution protocol use case. \textbf{(a)} In the three-user network, User 1 shares  entangled links with both other users, while Users 0 and 2 each have one virtual neighbor. An upcoming swap at User 1 will allow all users to reach their maximum of two virtual neighbors.  \textbf{(b)-(c)} illustrate larger networks. In all cases, the number of qubits per node is set by multiplying the node degree by $r=5$. (Note: link lengths are visually scaled for clarity and do not represent actual distances between nodes.)
} 
    \label{fig:cd-randomtree}
\end{figure}

Our final use case explores protocols for continuous distribution of entanglement in quantum networks. 
Unlike on-demand protocols that require specific scheduling policies to coordinate node operations based on user demands, continuous-distribution protocols distribute entanglement nonstop throughout the network\cite{chakraborty2019distributed}. This allows for immediate use of entanglement by any node as needed, supporting applications that continuously operate and consume entanglement in the background. In particular, we consider the continuous-distribution protocol introduced in ref.\cite{inesta2023performance}, and we aim to find parameter values that optimize its performance. This protocol operates in discrete time. In each time slot, all physically neighboring pairs of nodes in the network (i.e., nodes that share a physical communication channel, such as optical fiber) attempt entanglement generation simultaneously, where each attempt succeeds with probability $p_{\text{gen}}$. For simplicity, we assume that all link lengths are the same.
A pair of nodes can successfully generate entanglement in successive time slots and hold multiple entangled links, limited only by the number of qubits available at the nodes (see Figure \ref{fig:cd-randomtree}). Each node $i$ holds a number of memory qubits proportional to the number of physical neighbors: $r \cdot d_i$, where $r\in \mathbb{N}$ is a parameter specified in the protocol and $d_i$ is the degree of node $i$.

Entanglement between non-neighboring nodes $j$ and $k$ can be generated via entanglement swapping at a node $i$ if entangled links $i-j$ and $i-k$ already exist.
In each time slot, each node $i$ randomly chooses two links $i-j$ and $i-k$, and performs a swap operation on them with probability $q_{\mathrm{swap},i}\in [0,1]$ (otherwise, the swap is not performed and the links stay as they are). A successful swap consumes both links and produces a $j-k$ link. We assume that swaps succeed deterministically. As every swap decreases the quality of the entanglement\cite{azuma2023quantum}, limiting the number of consecutive swaps presents a practical measure to prevent excessive decoherence.
In the considered protocol, the number of adjacent swaps is limited to involve a maximum of $M$ physical links. Furthermore, as the quality of an entangled link generally worsens with time \cite{azuma2023quantum, khatri2019practical}, links that exist for longer than some cutoff time $t_{\text{cut}}$ are removed to exclude low-quality entanglement from the network.
Lastly, we assume that there is an application running on the network that consumes the entangled links generated by the protocol. Specifically, pairs of nodes consume an existing entangled link at each time slot with probability $p_{\text{cons}}$. We simulate this protocol over $T_{sim}$ time slots, where $T_{sim}$ is a number of discrete time units.

The utility in this use case quantifies the ability of a user to run background applications. This is determined by the number of readily available entangled links to other nodes in the network. Nodes connected by at least one entangled link are termed \textit{virtual neighbors}. The objective therefore is to maximize the number of virtual neighbors for all users, which can be accomplished by tuning the swap probability $q_{\mathrm{swap},i}$ at each node $i$. We use the vector $\mathbf{q_{\mathrm{swap}}} = [q_{\mathrm{swap},0}, q_{\mathrm{swap},1},\dots, q_{\mathrm{swap},N-1}]$  to represent all $N$ nodes' swap probabilities.

In contrast to previous use cases, our analysis of continuous-distribution protocols goes beyond aggregating objectives: it includes examining the individual user's goal to increase their own number of virtual neighbors. This approach is based on the Pareto frontier, a method of assessing multi-objective optimality, formally defined in e.g., ref.\cite{inesta2023performance}. As outlined in Section~\ref{sec:workflow}, the surrogate optimization process naturally collects instances of $\{\mathbf{q_{\mathrm{swap}}}\}$ and their simulation outcomes (objective values) as training data. We will refer to the former as the  \textit{collected set} $\mathcal{S}$ of solutions. From this data, we can find the dominating set $\mathcal{S}_\mathrm{dom}\subseteq\mathcal{S}$, which functions as an empirical estimate of the Pareto optimal set. It is important to note that this analysis of the collected solutions is not exclusive to this use case but could be applied to the other presented use cases as well. For clarity and illustrative purposes, however, we include this analysis only as part of this last section. 

\begin{definition}[Dominating Set $\mathcal{S}_\mathrm{dom}$]
Let $\mathcal{S}$ represent the set of all solutions collected during the surrogate optimization process, and let $\mathcal{S}_{\mathrm{dom}}$ denote the subset of these solutions that form the dominating set. A solution $\mathbf{q^*_{\mathrm{swap}}} \in \mathcal{S}$ belongs to $\mathcal{S}_{\mathrm{dom}}$ if and only if there does not exist another solution $\mathbf{q_{\mathrm{swap}}} \in \mathcal{S}$ such that $\mathbf{q_{\mathrm{swap}}}$ dominates $\mathbf{q^*_{\mathrm{swap}}}$. A solution $\mathbf{q_{\mathrm{swap}}}$ dominates $\mathbf{q^*_{\mathrm{swap}}}$ if:
\begin{equation}
\forall i \in \{1, \ldots, m\}, U^{(i)}(\mathbf{q_{\mathrm{swap}}}) \geq U^{(i)}(\mathbf{q^*_{\mathrm{swap}}}) \text{ and } \exists j \in \{1, \ldots, m\} : U^{(j)}(\mathbf{q_{\mathrm{swap}}}) > U^{(j)}(\mathbf{\mathbf{q^*_{\mathrm{swap}}}}).
\end{equation}
Thus, each solution in $\mathcal{S}_{\mathrm{dom}}$ is non-dominated with respect to the objectives $\{U^{(1)}, \ldots, U^{(m)}\}$, and any solution in $\mathcal{S} \setminus \mathcal{S}_{\mathrm{dom}}$ is dominated by at least one solution in $\mathcal{S}_{\mathrm{dom}}$.
\end{definition}
 For an intuitive explanation, see Supplementary Figure 3. In our approach, $\Bar{U}^{(i)}(f, \mathbf{q_{\mathrm{swap}}})$ denotes the average number of virtual neighbors per user $i$, which is generated by the simulation function $f(\mathbf{q_{\mathrm{swap}}})$ of ref.~\cite{inesta2023performance}.

First, we investigate a simple three-node quantum network, where all three nodes are users who are able to swap entanglement. Then, we extend our study to two larger asymmetric topologies involving $20$ and $100$ nodes, respectively. All three layouts are depicted in Figure \ref{fig:cd-randomtree}. In the two larger networks, all nodes are again able to swap entanglement, but only the leaf nodes (nodes with one physical neighbor) are considered \textit{users}. We explore the different setups by analyzing the dominating set of the collected solutions. However, we will also demonstrate that as the number of objective functions (i.e., the number of users) increases, a larger proportion of solutions is non-dominated, which is a well-known consequence of multi-objective optimization in high dimensional spaces\cite{kukkonen2007ranking}. Lastly, we conduct the already familiar comparison of aggregated objectives (sum over all users) found by surrogate optimization to the reference methods in the largest network topology. In our experiments we simulate continuous distribution protocols in quantum networks using parameter settings based on the previous study\cite{inesta2023performance}: $p_{\text{gen}} = 0.9$, $T_{\text{sim}} = 1000$, $r=5$, $M=10$, $p_{\text{cons}} = p_{\text{gen}}/4$, $t_{\text{cut}} = 28$ time units.

\subsubsection*{Three-User Network}
Using our surrogate framework, we simulate a three-user quantum network that implements the continuous-distribution protocol from ref.\cite{inesta2023performance} (see Figure \ref{fig:cd-randomtree}
 \textbf{a}). Post-optimization, we analyze the solutions and compute the dominating set\cite{kung1975finding}. We aim to optimize the network's virtual connectivity by tuning swap parameters ${q_{swap,i}}$ for each user $i$, where $q_{swap,1}=q_{swap,2}$ reflects the network's symmetry. As expected, the parameter ranges in the dominating solutions reflect the network configurations with overall largest number of virtual neighbors computed by the model (see Supplementary Figure 4). We observe a wide-spread distribution at the leaf nodes while the swap probabilities at the central node ($q_{\mathrm{swap},0}$) are concentrated around a low swap probability of 0.2. These findings can be explained as follows: entanglement swapping at Users 1 and 2 only establishes the virtual connections that already can be generated through direct entanglement generation at the physical links. This diminishes the impact of such swaps, and thus the swap parameters $q_{swap,1/2}$ are more widely distributed.  Entanglement swapping is indeed only crucial for the entangled link shared between Users 1 and 2 (as there is no physical link). Thus, given that swapping is necessary for only a small portion of virtual connectivity in this setting, a low swap probability at User 0 proves effective. With this analysis we demonstrate that even in small networks, the dominating set offers non-trivial insights into the performance of continuous-distribution protocols across multiple objectives.
 
\subsubsection*{Random Tree Networks}
 In the three-user setup, we could explain parameter performance by reasoning about the interactions between the three nodes. When scaling to larger networks, the number of interactions increases, but our objective remains the same. As in the previous setup, we evaluate the dominating set of solutions after implementing our surrogate optimization workflow.
In comparison, while the three-node network had $3.5$\% solutions dominating, the 20-node network in Figure \ref{fig:cd-randomtree}\textbf{b} has $121$ of $1000$ ($12$\%) solutions in $\mathcal{S}_{\mathrm{dom}}$. In addition to the standard deviation, we evaluate variation in a collected swap parameter distribution using the Kolmogorov-Smirnov (KS) statistic~\cite{lilliefors1967kolmogorov}, which measures the distance between two distributions. 

\begin{figure}[ht!]
    \centering
\begin{minipage}{0.45\linewidth}
\includegraphics[width=\textwidth]{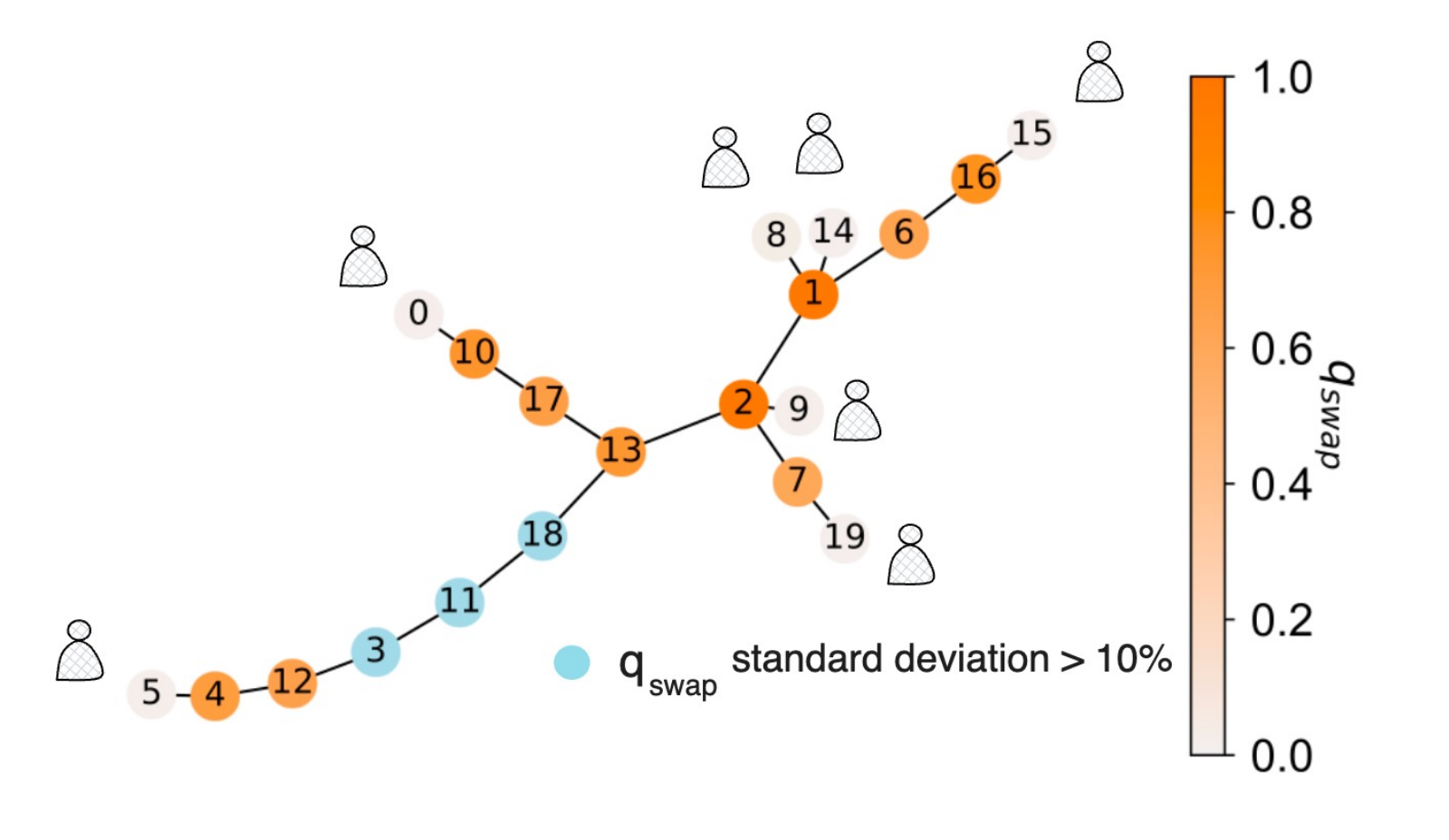}
\subcaption{\centering}
\end{minipage}
\begin{minipage}{0.45\linewidth}
\includegraphics[width=\textwidth]{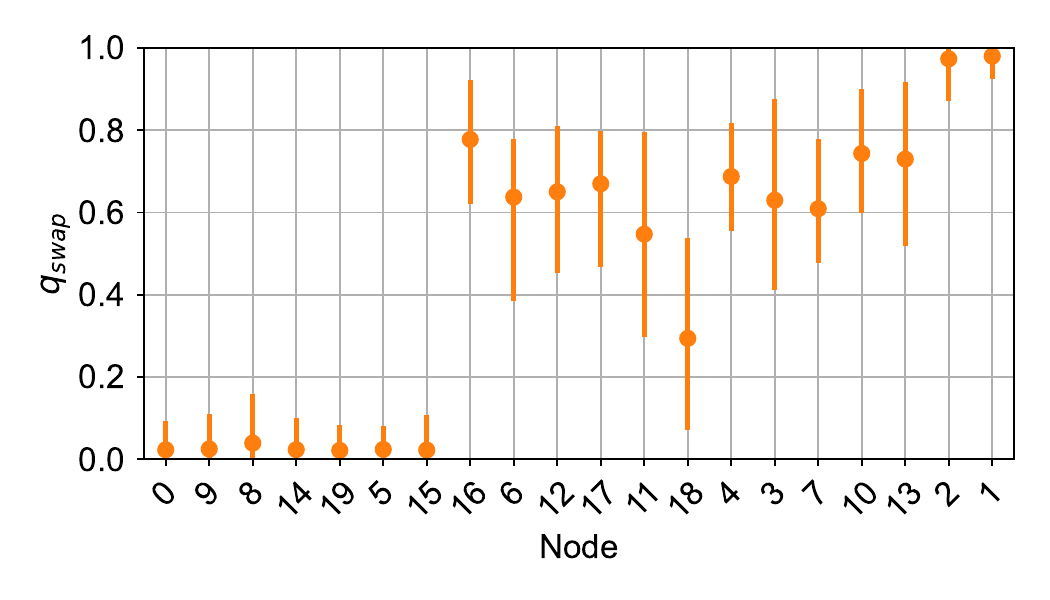}
\subcaption{\centering}
\end{minipage}
    \caption{\textbf{Swap probabilities of dominating solutions $\mathcal{S}_{\mathrm{dom}}$.}  \textbf{(a)} Most swap probability parameters exhibit a clearly biased distribution with standard deviation $< 0.1$, where the orange color on the color scale corresponds to the mean of 121 parameter values present in the dominating set of solutions. Three parameters (marked in blue) exhibit a wide-spread distribution, making the mean value less representative. \textbf{(b)} Median and a 95\% interval -- ranging from the 2.5 to the 97.5 percentiles -- of swap probabilities across nodes ordered by node degree. The distributions reflect the network's connectivity, where leaf nodes have a clear preference of swapping less frequently, while the nodes with higher vertex degrees exhibit higher swapping rates. Execution parameters used in the surrogate workflow: $T=100$ optimization cycles, $d=4$, $n=100$.}
    \label{fig:cd-randomtree20-pareto}
\end{figure}

Figure \ref{fig:cd-randomtree20-pareto} shows the swap probability distributions over solutions in $\mathcal{S}_{\mathrm{dom}}$ obtained from our surrogate workflow. Swap probability parameters at nodes 3, 11, and 18 show a larger standard deviation ($>0.1$) compared to other nodes. Additionally, these parameter values more closely align with a uniform distribution than with a normal distribution according to the evaluated KS measure. These metrics indicate that there is no definitive preference for swap parameter settings at these nodes. However, it is noteworthy that all other nodes exhibit more concentrated distributions: user nodes typically swap with a low probability (below $0.1$), whereas the nodes with the highest degree (nodes 1 and 2) engage in swaps almost every time slot (with probabilities exceeding $0.95$). These trends are anticipated, as users aim to maximize their virtual neighbors, and low swap probabilities help them accomplish this task. Conversely, nodes in the center should utilize their high connectivity to fulfill the users' objectives. This raises the question of why users should be allowed to swap at all, given that swapping could undermine their objectives. The rationale for permitting swaps at user nodes, albeit infrequently, is that a potential positive impact cannot be a-priori excluded. For example, a user near a highly connected hub (such as User 8) could enhance the outcomes for less ideally positioned users (e.g., Users 15 or 5) through occasional swapping, without loss of its own virtual neighbors on average.

Next, we examine the largest network in this study, depicted in Figure \ref{fig:cd-randomtree}\textbf{c}, involving 100 configurable swap probabilities. We apply our workflow under three different time constraints -- 1, 5, and 10 h -- yielding an average of 108, 590, and 950 solutions, respectively. Remarkably, 100\% of the solutions from the two shorter durations and 97\% from the longest duration were found to be non-dominated, reflecting the already anticipated \emph{curse of dimensionality}~\cite{kukkonen2007ranking}. All parameters in the dominating sets are closer to uniform than to normal distributions according to the KS measure, with a standard deviation greater than $0.1$, meaning that we should refrain from drawing generalized conclusions in this scenario.

\begin{figure}
    \centering
    \includegraphics[width=0.4\textwidth]{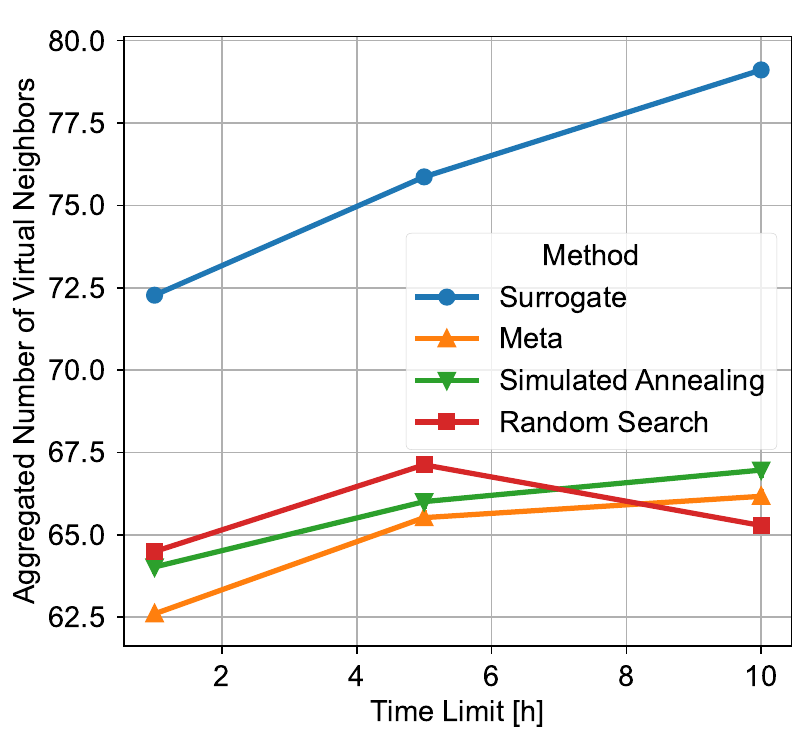}
    \caption{\textbf{Aggregated number of virtual neighbors for 100-node network and best found solution.} The best of ten solutions (of ten independent repetitions conducted, see \ref{sec:methods-experimentalsetup}) according to the objective function is selected. and a subsequent
simulation involving  runs. We find that all but Meta benefit from enlarged time constraints. Each point in the graph plot shown presents the mean of $n_{exec}=10^3$ simulation runs (with standard errors smaller than marker size). Execution parameters used in the surrogate workflow: $T=[1, 5, 10]$ h, $d=4$, $n=20$.}
    \label{fig:cd-randomtree-results}
\end{figure}

In addition to analysing the dominating set, we compare the aggregated number of virtual neighbors (sum over all objectives) to the reference methods (Section \ref{sec:methods-baselines}) under the same time constraints. We display the optimization results for all algorithms in Figure \ref{fig:cd-randomtree-results}. Remarkably, both Simulated Annealing and Meta can only outperform random search in the largest time limit, while our workflow performs significantly better in all scenarios, 
by achieving as much as 18\% more virtual neighbors in the largest time limit.
Bayesian optimization stays consistently below Simulated Annealing; this is likely because the 100 variables present in this network are far beyond the recommended limit of 20\cite{frazier2018tutorial}. Conversely, the models used by our surrogate optimizer are evidently not limited by the large number of variables, showing significant improvement with increased time limit. Note that even though random search identifies a solution within a ten-hour limit -- using $n=20$ simulation runs -- that exhibits more virtual neighbors than the solution found in the five-hour limit, the use of $n_{exec}=10^3$ simulation runs of the result depicted in Figure \ref{fig:cd-randomtree-results}  provides a more accurate estimate of the average, albeit a smaller one.

To conclude this section, we summarize the most relevant findings: across different quantum network sizes, the collected dominating set can provide relevant insights to the behaviour of parameters in continuous-distribution protocols when multiple objectives are to be met. Even for a 20-node setup, swap probability distributions are concentrated at some nodes, which can guide the configuration of these parameters in a protocol. However, the ratio of non-dominated to dominated solutions increases significantly with the number of objectives; in the largest network studied, this leads to all parameter distributions being closer to a uniform than to a normal distribution. Although the dominating set provided limited insights in the largest setup, we could still find a promising solution to the single-objective case, i.e., maximizing the sum over all users' virtual neighbors. The latter comprises 79 virtual neighbors on average, which is 18\% and 20\% higher than the best solutions found by Simulated Annealing and Bayesian optimization, respectively.

\section{Related Work} \label{sec:relwork}
In addition to the work discussed in the above scenarios, relevant prior studies were carried out by Ferreira da Silva et al.~\cite{da2021optimizing}, Wallnh\"ofer et al. ~\cite{Wallnöfer_2020},  Khatri et al.~\cite{khatri2021policies} and Haldar et al.~\cite{Haldar2024Fast}. Ref.~\cite{da2021optimizing} utilized genetic algorithms combined with simulation of repeater chains in NetSquid. In contrast to genetic algorithms, our approach utilized simple machine learning models to save on extensive evaluation of the expensive objective function, which helped to traverse the search space more efficiently. Further, ref.~\cite{da2021optimizing} focused on optimizing for minimum requirements to satisfy target benchmarks, while we investigated diverse utility functions based on different use cases.  Similar to our approach, the other studies~\cite{Wallnöfer_2020, khatri2021policies, Haldar2024Fast} used machine learning techniques to find optimal quantum network protocols. These studies were based on the theoretical framework of decision processes. Ref.~\cite{Wallnöfer_2020} introduced for the first time learning agents for quantum networks, which by trial and error manipulate quantum states and thereby construct communication protocols. Ref.~\cite{khatri2021policies} and ref.~\cite{Haldar2024Fast} devised a mathematical framework to optimize link-level entanglement generation in general networks allowing for said learning agents to discover policies. In contrast to our approach, learning agents are able discover complete protocols (devising optimal sequences of actions), while surrogate optimization is limited to finding improved parameter settings \emph{within} protocols. However, our approach offers two distinct advantages: 1) It is applicable to enhancing hardware configurations, such as the allocation of memory qubits per node. In contrast, learning agents operate within a static architectural environment, limiting them to discover only quantum network protocols; 2) We employ detailed quantum network simulations, in contrast to the purely mathematical or overly simplified models used in other studies. This allows us to address complex network topologies beyond, e.g., simple linear repeater chains typically considered in prior work.

\section{Methods} \label{sec:methods}
\subsection{Baselines} \label{sec:methods-baselines}
In order to evaluate our approach, we utilize the following baselines: 1) \textit{Uniform random search}: this method chooses sets of parameter values uniformly at random from $X_{\text{conf}}$ to execute the simulation. We employ this method as our foundational baseline because its constraints are easily managed through time or iteration limits. Uniform sampling prevents being constrained to a specific area of the search domain, a limitation that might be encountered by a time- or iteration-limited exhaustive grid search. 2) \textit{Simulated Annealing}\cite{kirkpatrick1983optimization}: this conventional global optimization method starts by accepting less optimal solutions with high probability, thus enabling exploration of the search space and escaping from local optima using the so-called Metropolis criterion. We implemented the algorithm using the fast annealing schedule\cite{kochenderfer2019algorithms} and found objective values of various benchmark functions\cite{jamil2013literature} comparable to the L-BFGS-B method\cite{liu1989limited}.
3) \textit{Bayesian optimization API by Meta}: Bayesian optimization\cite{jones1998efficient} is a surrogate method based on Gaussian processes, which is mostly used to optimize unknown but continuous functions; empirical studies suggest optimal performance below 20 variables~\cite{frazier2018tutorial}. The Service/Loop API, developed by Meta, incorporates a Bayesian optimization algorithm~\cite{balandat2020botorch}, where we adhere to the default configuration settings recommended in their official documentation (\url{https://ax.dev/docs/bayesopt}).

\subsection{Simulation Experiments} \label{sec:methods-experimentalsetup}
For each use case scenario, we compared optimization methods by conducting ten independent repetitions, each with an allotted time limit of $T$ hours. To compute averages $\Bar{U}^{(i)}(f,.)$ we used $n=20$ simulation runs by default. We report on the distribution of the ten repetitions in Supplementary Note~6. From all repetitions, the solution that performed best according to the objective function is selected, and a subsequent simulation involving $n_{exec} = 10^3$ simulation runs were executed. This approach leads to statistically robust solutions, characterized by small standard errors (explicit values given in captions of respective result figures). In the scenarios, where we did not compare to reference methods, $T$ is a set number of optimization cycles and $n\ge100$. Note that in scenarios involving time constraints for comparison between reference methods, the outcomes are specific to our computing system. For instance, let us assume a simulation takes nine seconds to execute on our system but ten seconds on another computer. If we impose a time limit that permits ten executions on our system, the same time limit would only allow nine executions on the slower one, resulting in different outcomes. Since execution times depend on multiple aspects of a computing node (processor speed, RAM, operating system), to ensure exact reproducibility of output data, it is advisable to use a fixed number of iterations; we diverged from this to enable fair comparison between used optimization methods. Each optimization method was allocated ten cores on an Intel(R) Xeon(R) CPU E5-2620 v2 @ 2.10GHz and 64 GB of memory.

\FloatBarrier
\section*{Conclusion}
We introduced a surrogate-based optimization workflow that integrates detailed quantum network simulations. Through three distinct use cases, we demonstrated the workflow’s broad applicability to practical optimization problems and its effectiveness in enhancing the performance of both on-demand and continuous entanglement distribution protocols across various network topologies. Particularly, the application to a five-user quantum entanglement switch and a  metropolitan network performing purification protocols, showed our method's efficacy to handle complex and highly asymmetric quantum networking scenarios. Our workflow efficiently handled up to 100 network parameters and achieved approximate solutions that significantly outperformed optimization techniques such as Simulated Annealing (up to 18\%) and Bayesian optimization (up to 20\%) in tested scenarios. Moreover, finding the dominating set of collected solutions allowed us to derive insights from a multi-objective perspective. 

Due to the potentially large computational costs of numerical simulation, our approach is suitable for scenarios where the simulation models an analytically intractable problem, i.e., a problem scenario in which it is justified to invest additional time in simulation and optimization. Overall, this work introduced a scalable and effective optimization tool for the discovery of new beneficial quantum network configurations and demonstrated its contribution in practical use cases.

\section*{Acknowledgements}
This work is supported by QuTech NWO funding 2020–2024 Part I ‘Fundamental Research’, Project Number 601.QT.001-1, financed by the Dutch Research Council (NWO). We further acknowledge support from
\emph{ NWO QSC grant BGR2 17.269.} Á.G.I. acknowledges financial support from the Netherlands Organisation for Scientific Research (NWO/OCW), as part of the Frontiers of Nanoscience program. 

\section*{Data and Code availability}
Source data, as well as well as generated machine learning models and time profiling data is available at \url{https://doi.org/10.4121/a07a9e97-f34c-4e7f-9f68-1010bfb857d0}. The code used to generate reported data, as well as usage guidelines can be found in the Git repository \url{https://github.com/Luisenden/qnetsur} and corresponding documentation \url{https://qnetsur.readthedocs.io}.





\small
\bibliography{References}

\appendix
\renewcommand{\thesection}{Supplementary Note \arabic{section}}
\setcounter{section}{0}
\section{Exploration vs exploitation in surrogate-assisted search}
\begin{figure}[ht!]
    \centering
    \includegraphics[width=0.5\linewidth]{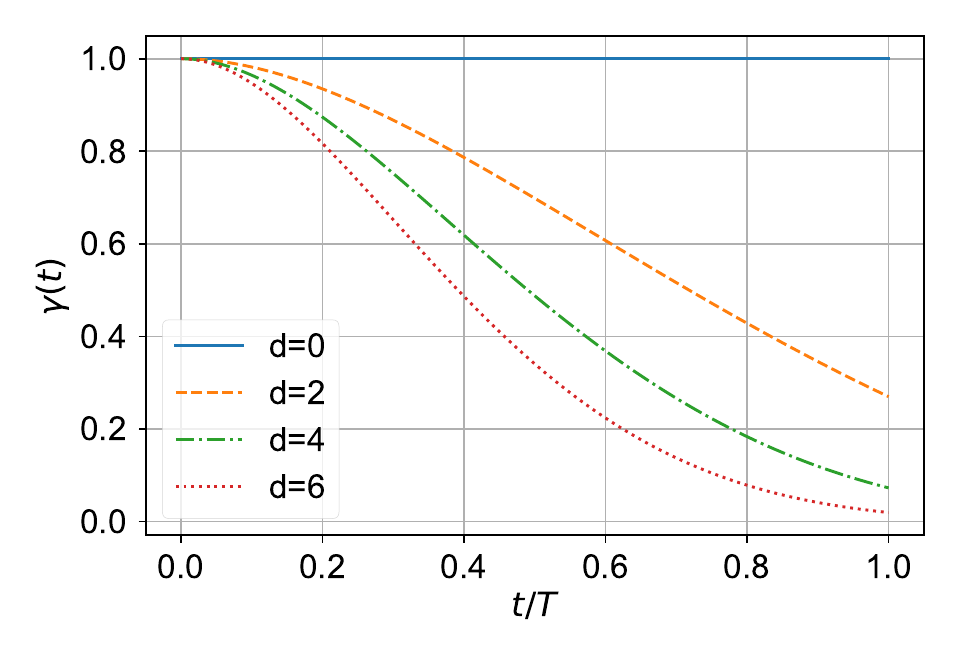}
    \caption{Transition function, where $t\in[0,T]$, using different exploitation degrees $d$.}
    \label{fig:transitionfunction}
\end{figure}

We apply the following exploration versus exploitation strategy in the introduced surrogate-assisted search: given a top configuration $\mathbf{s}^{\mathrm{top}} = \{x_1,\dots,x_N\}$, each optimization cycle $t$ decreases the standard deviation of a normal distribution $\mathcal{N}_{\mathrm{trunc}}(\mu_{p}, \sigma_{p}(t,d))$ around each parameter value $\mu_{p}\equiv x_p$. This way, the focus of discovering new, but less favorable configurations (exploration) gradually shifts towards refining known and well-performing configurations (exploitation). To achieve this, the distribution is truncated at $x_p^\mathrm{min}$ and $x_p^\mathrm{max}$ and the standard deviation narrows as $\sigma_p(t,d) = \gamma(t,d) (x_p^{\mathrm{max}}-x_p^\mathrm{min})/2$, with
\begin{align}
    \gamma(t,d) = (1-\ln^2(1+t/T))^d,\ \text{where } d\ge 1
\end{align} 
is a monotonically decreasing function for $t\in[0,T]$ (see proof below), $T$ is the maximum number of optimization cycles and $d$ is chosen according to desired degree of exploitation. For instance, $d=1$ leads to a minimum deviation of $ \sigma_p(t=T) = 0.52 \cdot (x_p^{\mathrm{max}}-x_p^\mathrm{min})/2$. In other words, by the end of the optimization process, the neighborhood, i.e., standard deviation, in which the models are mostly evaluated has been reduced to almost half its initial value; Figure \ref{fig:transitionfunction} shows four examples of the transition function, each with a different degree of exploitation.

\begin{proof}\label{proof:gamma}
A function is monotonically decreasing over an interval when its derivative remains negative throughout that interval. Thus our goal is to show that $\frac{\partial\gamma(t,d)}{\partial t} < 0$ for $0 \leq t \leq T$. 

\begin{equation}
\frac{\partial\gamma(t,d)}{\partial t} = -\frac{2d \ln(\frac{t+T}{T}) (1 - \ln^2(\frac{t+T}{T}))^{d-1}}{t+T}, \quad \text{for } t \in [0,T].
\end{equation}

Since $0 \leq \ln(\frac{t+T}{T}) < 1$, the term $(1 - \ln^2(\frac{t+T}{T}))^{d-1}$ must positive (recall that $d\geq 1$ by assumption). The negative sign in front of the fraction confirms that $\frac{\partial\gamma(t,d)}{\partial t}$ is less than zero and thus $\gamma(t,d)$ is monotonically decreasing on the interval $[0, T]$.
\end{proof}

\section{Acquisition process in surrogate-assisted search}
In each cycle $t$, the algorithm undertakes an acquisition process which involves two stages: model training and model evaluation. Initially, two machine learning models a -- Support Vector Regressor and a Decision Tree Regressor -- are trained independently. Their performance is then assessed through a five-fold cross-validation, using the mean absolute error as the metric on the current training data. The model demonstrating the lower error progresses to the acquisition phase.

During the acquisition phase, $N_t$ different configurations are sampled from the current normal distribution $\mathcal{N}_{\mathrm{trunc}}(\mu_p, \sigma_p(t,d))$ across all parameters $x_p\in{s^{top}_i}$, where ${s^{top}_i}$ is among the $l$ currently best performing parameter sets. Specifically, for each $x_p \sim \mathcal{N}_{\mathrm{trunc}}(\mu_p=x_p, \sigma_p(t,d))$ we retrieve $N_t$ values, generating $N_t$ sample sets $\{s^{\mathrm{eval}}_1,\dots, s^{\mathrm{eval}}_{N_t}\}$ to be evaluated. The number of samples $N_t = N(t)$ drawn increases with the number of elapsed cycles according to the formula $N(t) = 10 + 10^4 \cdot \frac{t}{T}$, allowing more computational resources to be dedicated during later cycles as the models gain more insights about the true objective function. The machine learning model's predicted objective values for the sampled configurations $\{s^{\mathrm{eval}}_1,\dots, s^{\mathrm{eval}}_{N_t}\}$ are sorted, and the highest-rated set is passed to the simulation for execution. The resulting parameter set, along with its simulated objective outcome, is then added to the training set. 

\section{Quantum entanglement switch use case}
In this supplementary note, we describe the quantum entanglement switch (QES) modelled in NetSquid, along with the results obtained when the switch serves two users. In total the system is simulated over $T_{\mathrm{sim}}=5$ seconds generating user-server links.

\subsection{QES Model}
In this model, the users and server continuously attempt to generate entangled links with the switch node via the single-click entanglement generation scheme. Once established, these entangled links are stored in available memory qubits, acting as a memory buffer. The primary function of the switch is to facilitate connections between server and each of the users through entanglement swapping.
\begin{enumerate}[leftmargin=*]
    \item \textit{Link-level entanglement generation}: In the single-click scheme assuming high photon losses, the fidelity of the produced states is given by $F_l = 1-\alpha_l$, where $\alpha_l$ is the bright-state population, a tunable experimental parameter. In this scheme, a state of the form
$$
\rho = (1 - \alpha_l) \ketbra{\Psi^+}{\Psi^+} + \alpha_l \ketbra{\uparrow\uparrow}{\uparrow\uparrow}
$$
is generated with probability
$$
p_{\text{gen,l}} = 2\eta_l \alpha_l,
$$
where $\ket{\Psi^+}$ is a Bell state orthogonal to the product state $\ket{\uparrow\uparrow}$, e.g., $\ket{\Psi^+} = \frac{1}{\sqrt{2}} (\ket{\uparrow\downarrow} + \ket{\downarrow\uparrow})$; and  $\eta_l$ is the transmissivity of link $l$ with length $L_l$, given by
$$
\eta_l = 10^{-0.1 \beta L_l},
$$
where $\beta = 0.2$ dB/km is the fiber attenuation coefficient. As the generation probability scales with the bright-state population $\alpha_l$, the latter facilitates a trade-off between rate and fidelity. 
 In our model, we approximate this physical behaviour with a depolarizing error happening with probability $p_l = \frac{4}{3}\alpha_l$ to a maximally entangled state, resulting in a Werner state with fidelity $F=1-\alpha_l$. Specifically, a depolarizing error degrades a perfect Bell state $\rho_{Bell} = \ketbra{\Psi^+}{\Psi^+}$ with a probability $p_l$ to the quantum state
\begin{align}
        \rho_{depol} = (1-p_l)\rho_{Bell} + \frac{p_l}{4}\mathbb{I}_4.
    \end{align}
The above state is a Werner state $\rho_{w} =  w\cdot\rho_{Bell} + (1-w)\frac{\mathbb{I}_4}{4}$ with Werner parameter $w = 1-p_l$ and fidelity $F = \frac{3w+1}{4}$. Within the QES simulation, the time interval between successful attempts follows an exponential distribution Exp($1/r$), with $r = \frac{p_{gen,l}}{T_{\mathrm{attempt}}}$, where $T_{\mathrm{attempt}} = 10^{-3}$s denotes the attempt-repetition time.
\item \textit{Memory Buffer}. When a new entangled state is generated, the state is shared by the so-called communication qubits -- one at the node (user/server) and one at the switch side -- with the purpose to host entangled states during the generation process. After successful generation, the entangled state is transferred (without any time loss) along with a dedicated timestamp to free memory qubits, one at the node and one at the switch.  Once in the buffer, the quantum states are assumed to be perfectly shielded, i.e., no noise is impairing the states. Should the buffer be full, the oldest state is discarded and the memory receives the fresh link. In our model, the users and server have 20 quantum memories at their disposal.
\item \textit{Entanglement swap.} The switch executes a Bell state measurement on the user-switch, and server-switch links to generate an end-to-end entangled link between user and server. By default, the user who holds the oldest link is chosen for the swap first (first come first served principle). Swaps are assumed to succeed deterministically with perfect gate operations. Further, the switch can only swap one link-pair at a time.  After a swap, the state's fidelity is recorded, and the state immediately discarded.
\end{enumerate}

\subsection{QES serving two users}
\begin{figure}[ht!]
 \centering
\begin{minipage}[b]{.5\textwidth}
    \centering
    \includegraphics[width=\textwidth]{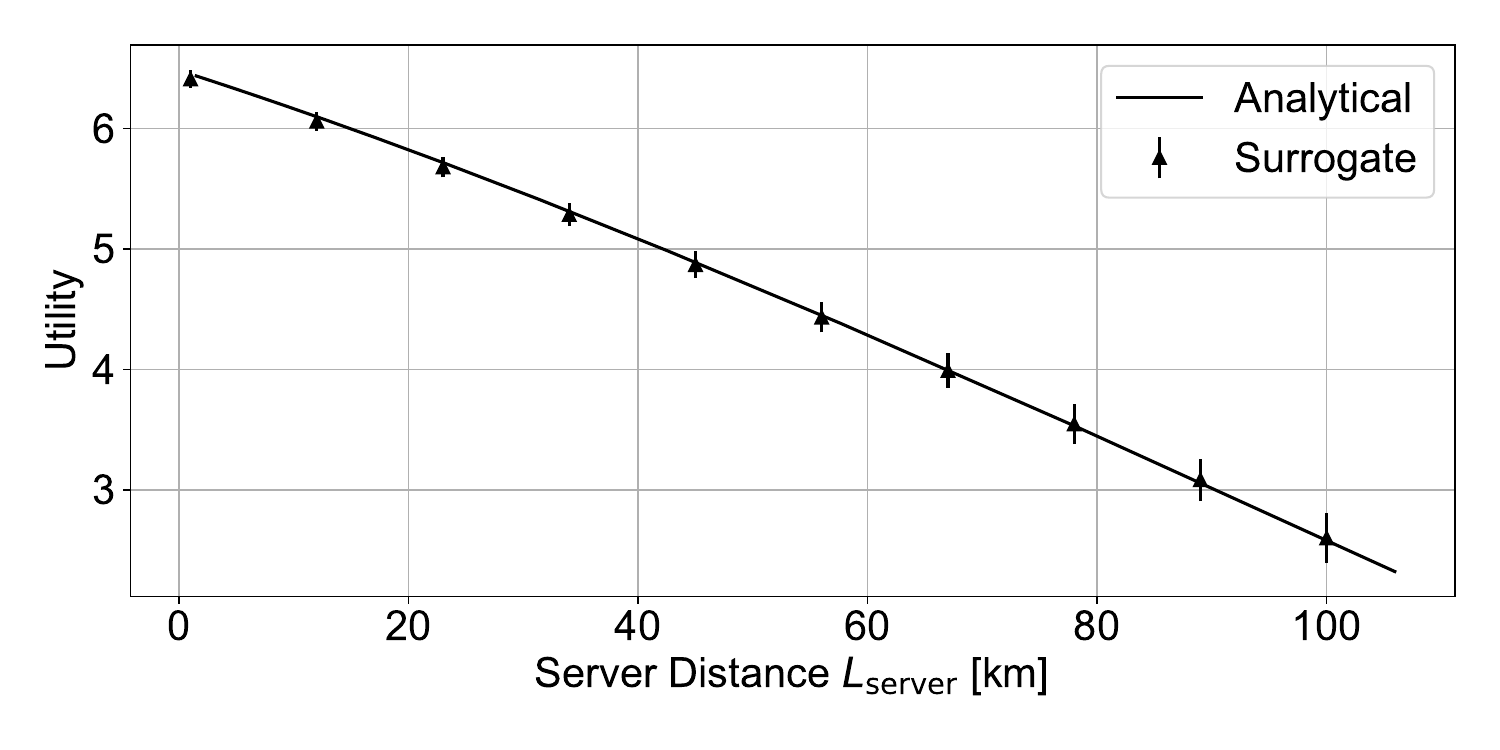}
\subcaption{\centering Utility based on distillable entanglement} \label{subfig:uscase1-example1-a}
    \end{minipage}
\begin{minipage}[b]{.35\textwidth}
  \centering\includegraphics[width=\linewidth]{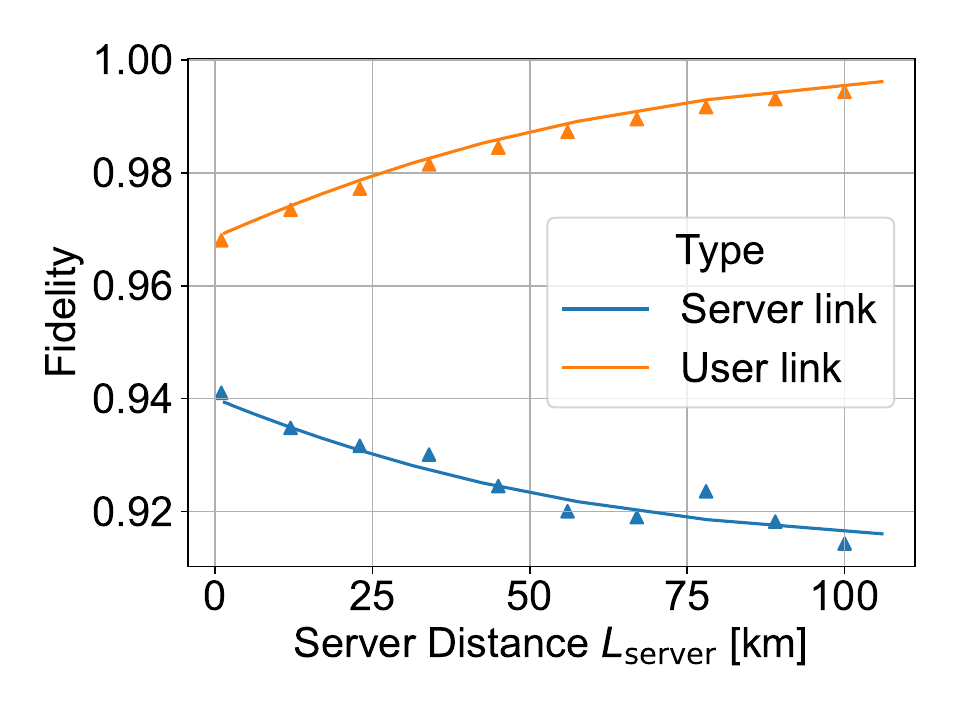}
  \subcaption{\centering Link-level fidelities $F_l = 1-\alpha_l$} \label{subfig:uscase1-example1-b}
\end{minipage} \\
\begin{minipage}[b]{.35\textwidth}
  \centering
  \includegraphics[width=\linewidth]{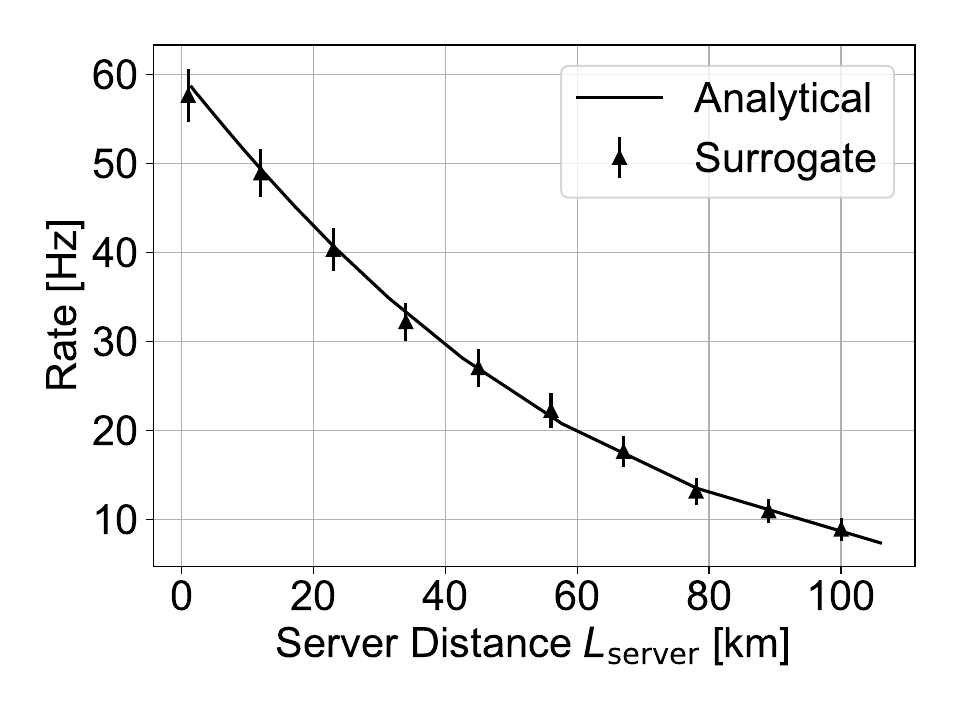}
    \subcaption{\centering End-to-end rates}\label{subfig:uscase1-example1-c}
\end{minipage}
\begin{minipage}[b]{.35\textwidth}
  \centering
  \includegraphics[width=\linewidth]{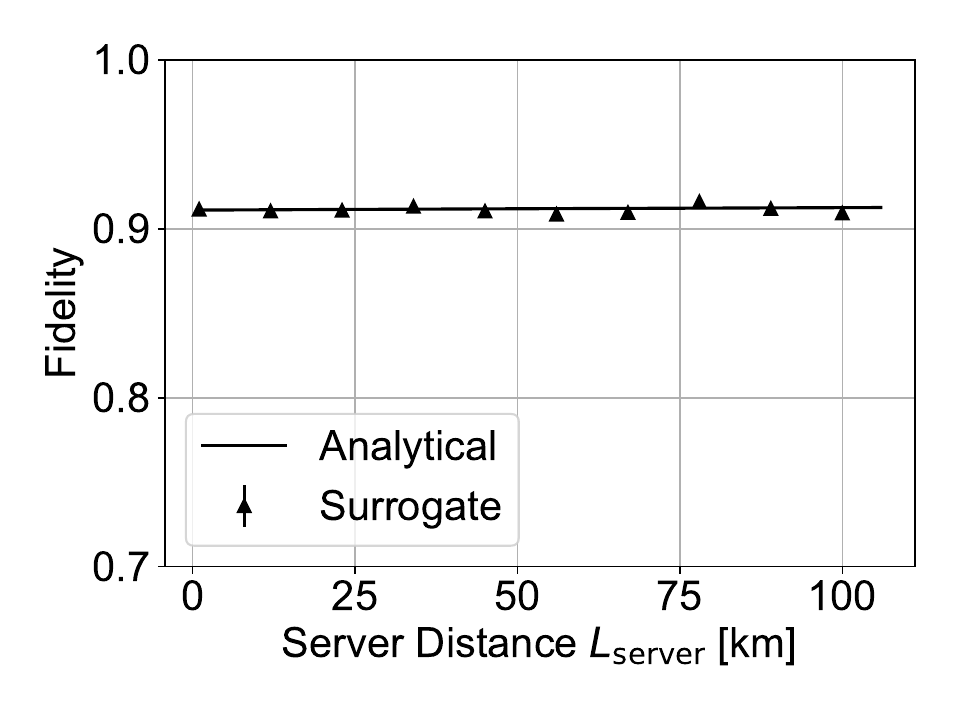}
  \subcaption{\centering End-to-end fidelities}\label{subfig:uscase1-example1-d}
\end{minipage}
\caption{Utility maximization over bright-state parameters for a QES serving two users and a server. Triangular markers present outcomes found via surrogate-assisted search, while the analytical model is given by solid lines. (a)-(b) Found aggregate utility and link-level fidelities $F_l = 1 - \alpha_l$. (c)-(d) Rates and fidelities of produced user-server links. Each marker in (a), (c), and (d) indicates the mean and  standard deviation of $n=100$ runs conducted in one simulation execution. Parameters used in the surrogate workflow: exploitation degree $d=4$, $T=100$ optimization cycles.}
  \label{fig:uscase1-example1}
 \end{figure}

We apply our surrogate workflow to a simple QES serving two users located two km from the switch node, while varying the server location between 2 km and 100 km from the switch. We thereby recover a selection of results from the analytical study by ref~\cite{vardoyan2023quantum}.
Figure \ref{subfig:uscase1-example1-a} shows that the optimization outcomes we find closely mirror the analytical model: utility decreases linearly with the distance of the server-switch link due to the network's decreasing ability to produce entanglement with high quality at high rates. When it comes to link-level fidelities $F_l$, see Figure \ref{subfig:uscase1-example1-b}, the server link as expected sacrifices the quality of the entanglement in order to accommodate the rates demanded from both the user sides. With increasing distance of the server to the switch, this trend is intensified; the user links' quality increases in order to make up for the quality lost in the server link. While not surprisingly the fidelities in Figure \ref{subfig:uscase1-example1-d} do not depend on the botteneck link's length, the rate exponentially decreases, see Figure \ref{subfig:uscase1-example1-c}.
\FloatBarrier
\section{Metropolitan network use case}
Here, we detail the modeling of the metropolitan network that serves user requests and outline the memory distribution policies identified using surrogate optimization and reference methods.
\subsection{Metropolitan Network Model}
The network configuration in our study is based on a simulation utilizing the SeQUeNCe software package (Wu et al., 2021). Network performance is assessed by the capacity to fulfill entanglement requests between user pairs. The process involves a user selecting another user uniformly at random and placing an entanglement request. Should a request not meet the required fidelity, it is re-attempted until success. Each user sequentially submits and completes a request, restarting the cycle once all have participated. The simulation time is set to $T_{\mathrm{sim}}=20$ seconds. 

\subsubsection*{Underlying hardware of quantum network model}
Below, we present an overview of the quantum network model introduced by Wu et al. (2021) and list the relevant hardware parameters in Table \ref{tab:parameters}. For a complete description of the model, we refer the interested reader to the original work.
\begin{itemize}
    \item[1.] \emph{Communication channels:} Quantum communication over telecommunication fiber links come with a propagation delay $L/c$, where $L$ is the fiber length and $c$ represents the speed of light within the fiber. Loss rates are quantified by the formula $10^{-L \cdot \alpha_o / 10}$, where $\alpha_o$ indicates the attenuation rate per kilometer in dB/km. To ensure orderly photon transmission, the system employs time-division multiplexing (TDM) to prevent overlap.
    \item[2.] \emph{Midpoint station}: Each midpoint station is equipped with a single-photon detector with a set efficiency, resolution and count rate (modelled as Poisson process).
    \item[3.] \emph{Quantum Memories:} During the process of entanglement generation, pairs of quantum memories located at different nodes undergo cycles of excitation and relaxation (memory frequency). This process results in each memory emitting a photon with probability $\eta|\alpha|^2$, depending on its state $\alpha\ket{\uparrow} + \beta\ket{\downarrow}$ and its memory efficiency $\eta$. At the midpoint station, these photons are detected, and the duration of the entangled state being held in memory is determined by the assumed coherence time\cite{ranvcic2018coherence}, after which the entangled state is discarded by resetting the memory qubits. 
    \item[4.] \emph{Purification:} Fidelity $\in [0,1]$ is a measure of closeness of a given quantum state with respect to a reference (e.g., desired) state. The initial fidelity of stored states changes when these states are subject to swap or purification operations. The utilized BBPSSW protocol can probabilistically improve the fidelity of an entangled state by sacrificing another entangled state. First,  operations are performed locally on both node sides on pairs of qubits holding an entangled state. Then the target qubits are measured in the Z-basis. The protocol then assesses whether to keep or discard the qubit pairs based on the measurement results. When both nodes measure their respective qubits and achieve the same results, it signals a successful purification operation, improving the entanglement fidelity of the unmeasured qubits. Conversely, differing results indicate a failed purification attempt, and the pairs lose their entangled state. 
\end{itemize}
\begin{table}[ht!]
\centering
\begin{tabular}{l l}
\hline
\textbf{Parameter} & \textbf{Value} \\
\hline
Memory efficiency $\eta$ & 0.75 \\
Memory frequency & 20 kHz \\
Memory coherence time & 1.3 s \\
Memory fidelity & 0.991\\
Detector efficiency & 0.8 \\
Detector count rate & 50 MHz \\
Detector resolution & 100 ps \\
Attenuation ($\alpha_o$) & 0.2 dB/km \\
Channel TDM time frame & 20 ns \\
Gate fidelity & 0.99 \\
Swap success probability & 0.64 \\
Swap degradation & 0.99 \\
\hline
\end{tabular}
\caption{Parameter values used in metropolitan network simulation model.}
\label{tab:parameters}
\end{table}

\subsection{Memory distribution policies for metropolitan network}
Table \ref{tab:usecase2-solutions} depicts the memory allocations found by surrogate-assisted search and reference methods, alongside a simple uniform allocation, termed Even, and the approach introduced in Wu et al., 2021.

\begin{table}[ht]
    \centering
\begin{tabular}{lllllll}
\toprule
Node & Surrogate & Meta & Simulated Annealing & Random Search & Wu et. al, 2021 & Even \\
\midrule
NU & 24 & 26 & 32 & 24 & 25 & 50 \\
StarLight & 73 & 79 & 88 & 34 & 91 & 50 \\
UChicago PME & 65 & 47 & 55 & 46 & 67 & 50 \\
UChicago HC & 26 & 40 & 31 & 33 & 24 & 50 \\
Fermilab 1 & 60 & 54 & 63 & 86 & 67 & 50 \\
Fermilab 2 & 25 & 31 & 43 & 23 & 24 & 50 \\
Argonne 1 & 89 & 67 & 61 & 75 & 103 & 50 \\
Argonne 2 & 39 & 36 & 29 & 73 & 25 & 50 \\
Argonne 3 & 25 & 30 & 43 & 51 & 24 & 50 \\
\bottomrule
\end{tabular}
\caption{Number of memories allocated per user node.}
\label{tab:usecase2-solutions}
\end{table}

\section{Estimated Pareto front of a three-node quantum network using continuous entanglement distribution protocols}
\begin{figure}[ht!]
    \centering
    \includegraphics[width=0.4\linewidth]{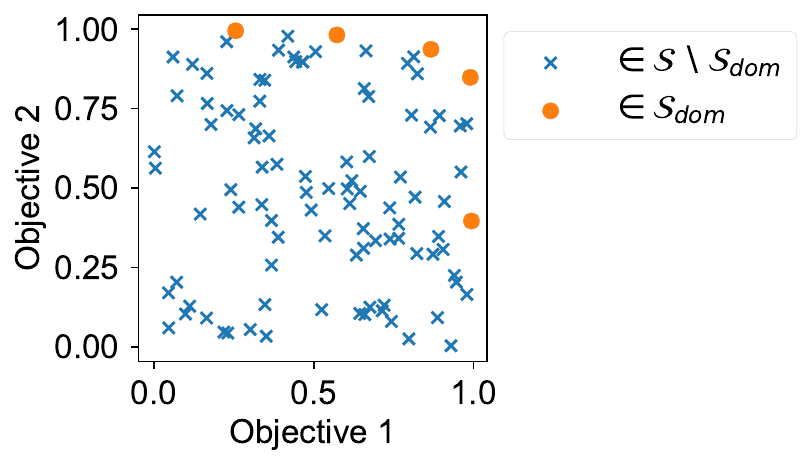}
    \caption{Estimated Pareto frontier $\mathcal{S}_{\text{dom}}$ in two maximization objectives (using uniform random data). A solution $\in \mathcal{S}_{\text{dom}}$ (orange) is not worse in any objective, and better in at least one objective than solutions $\in \mathcal{S}$ (blue).}
    \label{fig:simplepareto}
\end{figure}

\begin{figure}[ht!]
\begin{minipage}{.45\textwidth}
  \centering
  \includegraphics[width=\linewidth]{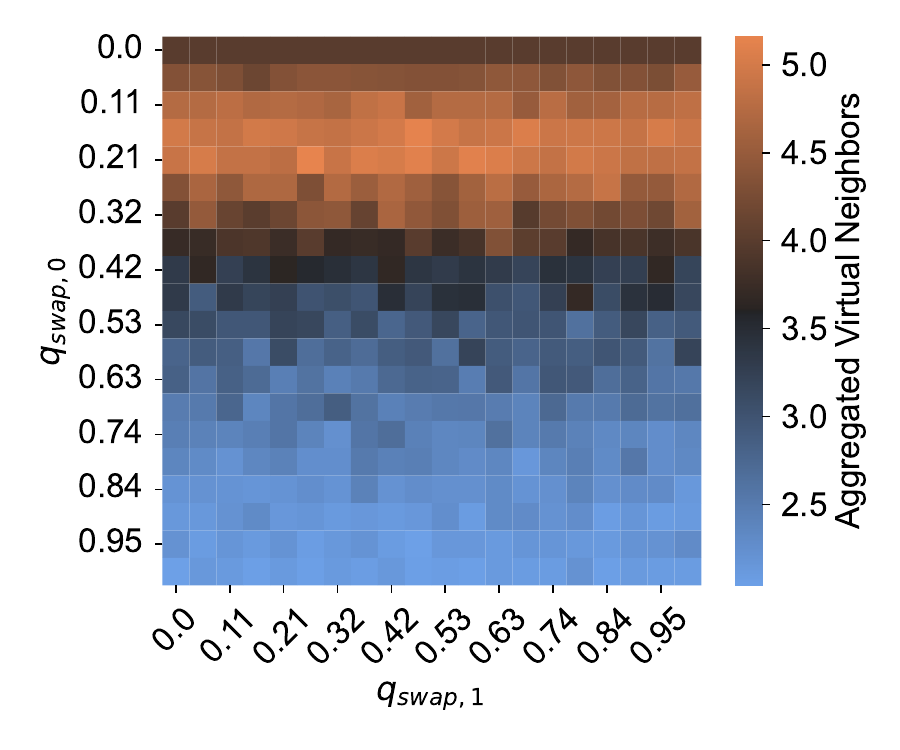}
\subcaption{\centering Aggregated number of neighbors for three-user network.}  
\end{minipage}%
\begin{minipage}{.45\textwidth}
  \centering
  \includegraphics[width=\linewidth]{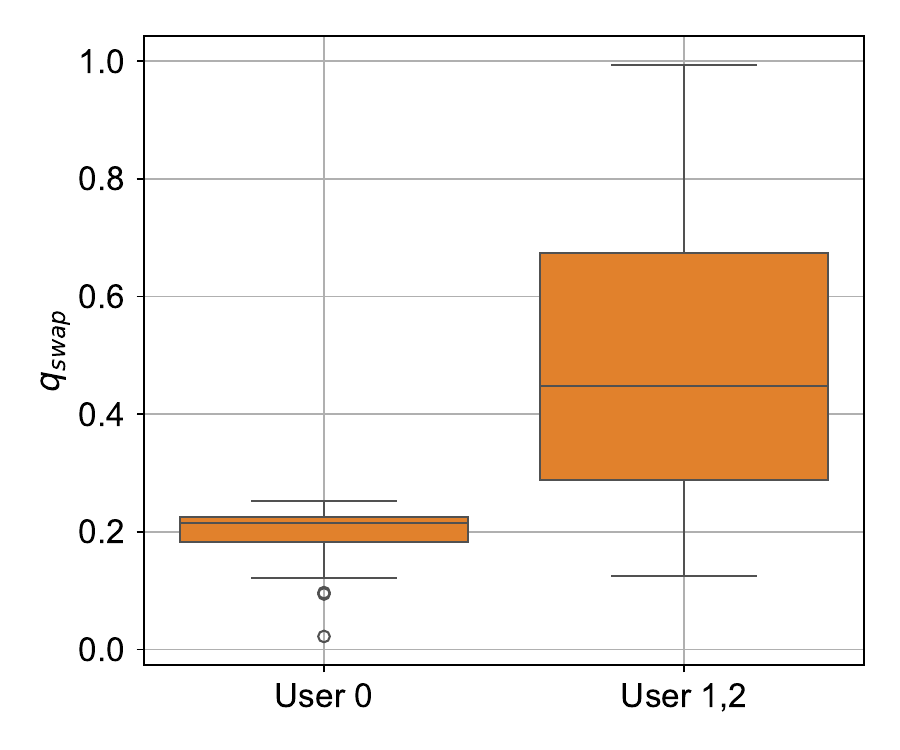}
\subcaption{\centering Distribution of parameter values in $\mathcal{S}_{\mathrm{dom}}$.}  
\end{minipage}
    \caption{(a) Aggregated number of virtual neighbors generated by the simulation for different swap probabilities $p_{\text{swap},1/2}$. Each grid square presents the mean of $n_{\mathrm{exec}}=10^3$ simulation runs (with a standard error below 0.01). The largest amount (light orange) of virtual neighbors can be provided by the network, when the protocol swap probabilities are small for User 0; the swap probabilities for the Users 1 and 2 have less significant impact. (b)  Boxplots present each a distributions of 35 values in the collected  solution set $\mathcal{S}_{\text{dom}}$ out of a over thousand (3.5\%) collected solutions.  We use the following parameter settings in our surrogate workflow: $T=100$ optimization cycles, $d=4$, $n=1000$.}
    \label{fig:cd-21tree-pareto}
\end{figure}

The surrogate optimization process collects instances of network-parameter values (i.e., parameter sets), which we append to the \textit{collected set} $\mathcal{S}$. From this data, we find the dominating set $\mathcal{S}_\mathrm{dom}\subseteq\mathcal{S}$, which functions as an empirical estimate of the Pareto optimal set.
Figure \ref{fig:simplepareto} shows an exemplary sample of random data and its collected dominating set $\mathcal{S}_\mathrm{dom}$ in two objectives. In Figure \ref{fig:cd-21tree-pareto} we show the aggregated number of virtual neighbors when executing the simulation over the whole parameter domain alongside the probability values of the dominating solutions $S_{\text{dom}}$. The two swap parameters are clearly distinguishable in $\mathcal{S}_{\text{dom}}$: While the swap-probability values of Users 1 and 2 are spread out over the whole range, the swap parameter values of User 0 are concentrated around 0.2 with a relatively small standard deviation (of 0.05). These results can be interpreted as follows: The majority of virtual neighbors are established directly through entanglement generation at the physical links. Consequently, swapping becomes essential primarily for the entanglement shared between Users 1 and 2, who lack a direct physical connection. Since only one third of the entangled links are generated exclusively through swapping at User 0, maintaining a low swap probability for User 0 proves sufficient.

\FloatBarrier
\section{Distribution of found objectives and time profiling}
In this supplementary note, we provide an overview of the distribution of objective values obtained from our surrogate optimizer and reference methods across use cases. Additionally, we detail the time allocation of the surrogate optimizer across different tasks.  Table \ref{tab:stability} summarizes the distribution of run results, where a smaller sample size $n$ results in a broader distribution. For instance, in the Metropolitan Network use case, a small number of simulation evaluations $n=5$ yields a relatively high standard deviation of 10\%. In contrast, the standard deviation is below 3\% in the QES and continuous-protocols use cases. Table \ref{tab:profiling} details the time spent in the surrogate-assisted search. As expected simulation consumes most of the optimization time, accounting for 85\% up to 99\% of the total execution time $T$ across all use cases.

\begin{table}[ht]
    \centering
        \begin{tabularx}{\textwidth}{l|cc|cc|cc}
    \multicolumn{1}{c}{} & \multicolumn{2}{c}{\textbf{QES}} & \multicolumn{2}{c}{\textbf{Metropolitan}} & \multicolumn{2}{c}{\textbf{Continuous Protocols}}\\ \hline
    \multicolumn{1}{l}{\makecell{Utility\\ based on}} & \multicolumn{2}{c}{\makecell{Distillable\\Entanglment}} & \multicolumn{2}{c}{\makecell{Completed\\Requests}} & \multicolumn{2}{c}{\makecell{Virtual Neighbors\\ ($T=$ 1 $|$ 5 $|$ 10 h)}} \\
        \hline
         \textbf{Method}&  Mean  & Std (rel.\%) & Mean &  Std (rel.\%) & Mean  & Std (rel.\%) \\
        Surrogate& 11.1 & 0.3 (3)& 34.7 & 3.4 (10) & 70.1 $|$ 77.1 $|$ 80.1 & 2.0 (3) $|$ 1.7 (2) $|$ 1.3 (2)\\
        Meta& 11.3 & 0.1 (1)& 28.8 & 4.6 (16)& 63.9 $|$ 65.4 $|$ 65.8 & 0.7 (1) $|$ 1.1 (2) $|$ 0.9 (1)\\
        Simulated A.& 10.3 & 0.4 (4)& 24.0 & 8.0 (33)& 64.0 $|$ 65.8 $|$ 66.8 & 1.0 (2) $|$ 1.2 (2) $|$ 1.0 (2)\\
        Random S.& 10.0 & 0.5 (5)& 18.2& 9.2 (51)& 63.3 $|$ 64.7 $|$ 65.1 & 1.3 (2) $|$ 0.8 (1) $|$ 1.2 (2)\\
        \bottomrule
        \end{tabularx}
    \caption{Distribution of maximum empirical utility values. Values present mean and (relative) standard deviation of ten independently conducted optimization experiments per method across investigated use cases.}
    \label{tab:stability}
\end{table}

\begin{table}[ht]
    \centering
        \begin{tabular}{l|rrr}
    &QES ($T=30$ min) & Metropolitan ($T=25$ h) & Continuous Protocols ($T=$ 1 $|$ 5 $|$ 10 h) \\ \hline
       Simulation & 99.1 \% & 99.78 \% & 95.0 $|$ 87.2 $|$ 85.3 \% \\
       Training &  0.1 \% & 0.01 \% & 0.3 $|$ 1.4 $|$ 1.9 \%\\
       Acquisition & 0.6 \% & 0.2 \% & 4.5 $|$ 11.2 $|$ 12.7 \% \\
    \textcolor{gray}{Remaining} & \textcolor{gray}{0.2 \%}& \textcolor{gray}{0.01 \%}& \textcolor{gray}{0.2 $|$ 0.2 $|$ 0.1 \%} \\
    \# Cycles & 26.9  & 35.2& 16.0  $|$ 72.3 $|$ 112.6  \\
        \bottomrule
        \end{tabular}
    \caption{Time profiling of ten independently conducted surrogate-assisted optimization experiments. \textit{Simulation} measures execution time of the simulation function at each optimization step. \textit{Training} refers to the time taken to construct machine learning models, while \textit{Acquisition} denotes the time required to identify a promising set of next execution points using these models. \textit{Remaining} includes initial object instantiation and writing to output files. The number of cycles represents the average count of optimization cycles completed within the prescribed time limit $T$. Values present averages of ten runs per use case with a standard deviation below 1\% in time acquisition and 5\% in the optimization cycles.}
    \label{tab:profiling}
\end{table}

\end{document}